\documentclass[a4paper,10pt,twoside]{article}
\usepackage[cp1251]{inputenc}
\usepackage[english,russian]{babel}
\usepackage{amsmath}
\usepackage{amssymb}
\usepackage{amsthm}
\usepackage{graphicx}

\theoremstyle{definition}

\theoremstyle{remark}

\newcommand{\bk}{{\bf k}}

\newcommand{\bR}{{\bf R}}
\newcommand{\bS}{{\bf S}}
\newcommand{\bP}{{\bf P}}

\newcommand{\bL}{{\bf L}}

\newcommand{\bX}{{\bf X}}

\newcommand{\bx}{{\bf x}}
\newcommand{\bz}{{\bf z}}
\newcommand{\by}{{\bf y}}
\newcommand{\bp}{{\bf p}}
\newcommand{\bC}{{\bf C}}
\newcommand{\bq}{{\bf q}}

\newcommand{\bZ}{{\bf Z}}
\newcommand{\hk}{{\hat{\bf k}}}

\newcommand{\homega}{{\hat{\bf \omega}}}

\newcommand{\hx}{{\hat{\bf x}}}
\newcommand{\hy}{{\hat{\bf y}}}

\newcommand{\hp}{{\hat{\bf p}}}

\newcommand{\be}{\begin{equation}}
\newcommand{\ee}{\end{equation}}
\newcommand{\bay}{\begin{eqnarray}}
\newcommand{\eay}{\end{eqnarray}}
 \author{Budylin A.M., Koptelov Ya.Yu., Levin S.B., Sokolov S.V.}
 \title{On asymptotic structure of three-body scattering states for the scattering problem of charged quantum particles.
\footnote{Ya.K. thanks Russian National Foundation for support in the frames of the grant RNF 17-11-01003 }
}
\begin{document}
 \addtolength{\hoffset}{-3.1cm}
  \addtolength{\voffset}{-3cm}

\maketitle

{\bf Abstract}.
We study the quantum scattering problem of three three-dimensional charged particles 
involving pair potentials of Coulomb attraction in the framework of the diffraction approach.  
We present for the first time the quantitative description of the influence of 
Coulomb pair excitations unified contribution in many-particle reactions.

\section{Introduction}

The diffraction approach in few-body quantum scattering problems with slowly (in the Coulomb way) decreasing repulsion pair potentials was previously discussed in literature 
\cite{BMS1}-\cite{L1} together with other approaches used in many-body scattering problems. Nevertheless the most interesting from the physical point of view case of quantum scattering of few Coulomb clusters (the system being in the Coulomb bound state) for a number of reasons remains not yet studied in detail despite the serious efforts made since the middle of the last century in the direction. At that one of the main difficulties together with the slow decrease of pair potentials and therefore with the difficulty of the construction of an asymptotic bound condition uniform at all angular variables in configuration space remains the proper accounting of the infinite number of excitation levels in pairwise subsystems. From the mathematical point of view it is connected with the presence of the discrete spectrum accumulation point of the Schroedinger operator corresponding to the pairwise subsystems with a Coulomb attraction potential.

The problem is not so much in making quantum calculations of scattering characteristics in such few-body subsystems. It is evidently being made, for example, by limiting the number of the accounted by scattering bound states in the subsystems with attraction or by choosing a certain (big) set of basis functions well describing experimental data in a certain energy range. The serious problem resides in specifying a set of criteria able to regulate the accuracy of such calculations in different energy ranges (or in dependence of energy).

In a sense it is connected with defining a number of excitation levels in pairwise subsystems, to be taken into account to obtain a calculation of the scattering characteristics with the required accuracy in dependence of the system energy.

We will study the problem here from the point of view of the diffraction approach mentioned above. The first steps on that way were already made in \cite{BKL}.

\section{Model Specification. }

We will give a more accurate model specification. The initial configuration space of the system is $\bR^9$. Having stopped the center of mass motion, we arrive at a system in configuration space
$$
\Gamma=\{\bz:\, \bz\in\bR^9,\,\bz=\{\bz_1,\bz_2,\bz_3\},\ \bz_1+\bz_2+\bz_3=0\}.
$$
On $\Gamma$ there is a scalar product $\langle\bz,\bz'\rangle$,
induced by the scalar product in $\bR^9$. The system on $\Gamma$ is described by the equation
$$
H\Psi=\lambda\Psi,\ \ \ \Psi=\Psi(\bz)\in\bC,\ \ \bz\in\Gamma,\ \ \ \
H=-\Delta_\bz+V(\bz), \quad V(\bz) = v_1(\bx_1)+v_2(\bx_2)+v_3(\bx_3),\ \ \ \bx_j\in\bR^3.
$$
Here $\Delta_\bz$ -- is a Laplace operator on $\Gamma$,
$$
 \bx_1=\frac{1}{\sqrt{2}}(\bz_3-\bz_2),\ \  \bx_2=\frac{1}{\sqrt{2}}(\bz_1-\bz_3), \ \
  \bx_3=\frac{1}{\sqrt{2}}(\bz_2-\bz_1).
$$
It is clear that $\bx_1+\bx_2+\bx_3=0$. Introduce also  $\by_j=\sqrt{\frac{3}{2}}\bz_j$. It is easy to see that on $\Gamma$ the equation $\by_1+\by_2+\by_3=0$ is true, as well as а
$$
\bz^2=\langle\bz,\bz\rangle=\langle\bx_j,\bx_j\rangle+\langle\by_j,\by_j\rangle,\ \ \ j=1,2,3\ \ \ , \quad \Delta_\bz = \Delta_\bx + \Delta_\by.
$$
Together with  $\bz\in\Gamma,\ \bx,\by\in\bR^3$ we will consider dual variables, impulses $\bq \in \Gamma,\ \ \bk,\bp \in \bR^3$.
We will assume that
$$
v_l(\bx)=\frac{\alpha_l}{x},\ \ \ \alpha_l=\sqrt{2\mu_{ij}}Z_i Z_j,
$$
where $\{ijl\}$ -- is a fixed even permutation of the numbers  $1,2,3$,  $\mu_{ij}=\frac{m_im_j}{m_i+m_j}$ -- is a reduced mass in the particles pair with the indices ${i,j}$, $Z_i$ $Z_j$ -- are the charges of the $i$-th and $j$-th  particles.
A generalization is also possible for the case
$$
v_l(\bx)= \frac{\alpha_l}{x} + w_l(\bx),\ xw_l(\bx) \to 0,\ x \to \infty,\ \ \ l=1,2,3.
$$
It should be remembered that the particles masses will be assumed equal, while the charges will be considered arbitrary in absolute value and of different sign, for example $m_1=m_2=m_3=1,\ Z_1=1,\ Z_2=2,\ Z_3=-3$.

\section{BBK -Approximation.}

The plane wave by the asymptotic (by $z\rightarrow\infty$) description of the function $\Psi(\bz,\bq)$ outside the small angular vicinities of the domains $\sigma_{j} = \{\bz \in \Gamma,\ \bx_j = 0\}$, $\ j = 1,2,3$ (the domains hereafter to be called screens) must be replaced by the BBK-approximation
$\Psi^{BBK}(\bz,\bq)$. The approximation was studied in \cite{BBK}, see also \cite{MF},though it was used previously as well (see, for example, \cite{z1,z2}). It looks as follows:
\be
\Psi^{BBK}(\bz,\bq)\sim N_0 e^{i\langle\bz,\bq\rangle}D(\bx_1,\bk_1)D(\bx_2,\bk_2)D(\bx_3,\bk_3).
\label{bbk0}
\ee
Here
\be
D(\bx,\bk) = \Phi(-i\eta, 1, ixk - i<\bx,\bk>), \ \bx,\bk \in {\bR}^3,\ \ \eta = \frac{\alpha}{2k},
\label{phi-1}
\ee
$\Phi$ - is a confluent hypergeometric function, see \cite{Grad}. The constant $N_0=\mathop{\prod}\limits_{j=1}^3N_c^{(j)}$ is a product of the normalization constants of three two-body scattering states
$N_c^{(j)}=(2\pi)^{-\frac32}e^{-\frac{\pi\eta_j}{2}}\Gamma(1+i\eta_j)$. The variables $\bk_j,\ \bp_j;\ $ $j=1,2,3$
are respectively conjugate in a Fourier sense to the Jacobi coordinates  $\bx_j,\ \by_j$.

It is worth noting that the function
\be
\psi_c(\bx,\bk) = N_c e^{i<\bx,\bk>}D(\bx, \bk)
\label{pair}
\ee
satisfies the equation
\be
-\Delta_{\bx} \psi_c + \frac{\alpha}{x}\psi_c = k^2 \psi_c.
\label{pair-eq}
\ee
The solution $\psi_c$  is an accurate solution of the Coulomb potential quantum particle scattering problem.
Hereafter for the convenience we will use another shorthand notation
\be
\Phi_i\equiv \Phi(-i\eta_i, 1, ix_ik_i - i<\bx_i,\bk_i>),\ \ \ i=1,2,3.
\label{pair-1}
\ee

Note as well that the BBK-approximation is true only in the domain where both Jacobi coordinates corresponding to the chosen pairwise subsystem turn large. To be more exact,  \cite{BBK}, \cite{MF}, \cite{BL1}, \cite{L1} we introduce a domain
\be
\Omega_\mu=\mathop{\cup}\limits_{j=1}^3\Omega_j,
\label{e_mu}
\ee
\be
\Omega_j=\{(\bx_j,\by_j),\ \ y_j^\mu < x_j < y_j,\ \ \frac12<\mu<1,\ \ y_j\rightarrow\infty\}
\label{e_mu_j}
\ee
Note that the BBK presentation ceases to be true in the asymptotic (by $r=\sqrt{x_j^2+y_j^2}\rightarrow\infty$, $\ j=1,2,3$)domains of the configuration space, where the Jacobi coordinate $\bx_j,\ \ j=1,2,3$ turns finite.
The main result of the works \cite{BL3}, \cite{L1} is exactly in the construction of the BBK -approximation continuous extension into the asymptotic domains of the configuration space by the finite as well as small values of the Jacobi coordinate $\bx_j,\ \ j=1,2,3.$
 $\bx_j,\ \ j=1,2,3.$
Note once again that these results were obtained in an assumption of the identity of the charge sign of all the particles included into the system, i.e. only for the case of repulsive pair potentials. The case to be considered now is a much richer one, as well as a much more complicated one, as it supposes an availability/presence/existence in some pairwise subsystems (with a Coulomb attraction pair potential) of an infinite set of excitation levels or, to put it differently, of an infinite number of asymptotic scattering channels.

We start, just as in the case of repulsive pair potentials, again by using the ideas of an "almost separation of variables".

\section{Asymptotic Almost Separation of Variables.}

Note that in the asymptotic "parabolic" vicinity of each screen
$\sigma_j=\{(\bx_j,\by_j):x_j=0\},\ \ j=1,2,3$
the Schrodinger equipment allows a serious simplification, after which a separation of variables becomes possible. For example, the total potential
$$
V(\bz)= v_1(\bx_1) + v_2(\bx_2) + v_3(\bx_3),\ \ v_i(\bx_i)=\frac{\alpha_i}{x_i},\ \ i=1,2,3
$$
in the asymptotic vicinity of the screen $\sigma_1\ \ $ ($y >> 1$)   is simplified at the cost of the formulas
$$
\bx_2 = - \frac{\sqrt{3}}{2}\by -\frac{1}{2}\bx,\ \bx_3 = \frac{\sqrt{3}}{2}\by -\frac{1}{2}\bx,\
\bx = \bx_1,\ \by = \by_1.
$$
In the vicinity $\sigma_1$ by  $y >> 1,\  y >> x$ true is
\be
V=V_{sep}+O\left(\frac{x}{y^2}\right),\ \ V_{sep} = v_1(x_1) + v^{eff}(y_1),\ \
v^{eff} = \frac{\alpha_{eff}}{y_1},\ \ \ \alpha_{eff}=2(\alpha_2+\alpha_3)/\sqrt{3}.
\label{eff-1}
\ee
(Hereafter for simplicity we will omit the index 1 of the corresponding Jacobi coordinates, as the procedure itself of approximations correlation will be held as an example in the parabolic vicinity of the screen 1.)
Therefore the first summand in the expression (\ref{eff-1}) describes well the potential $V$ as long as the correction $O\left(\frac{x}{y^2}\right)$ decreases faster than the Coulomb potential. Thus we are interested in the asymptotic domain
\be
\Omega_1^+: \{x\leq y^\nu,\ \ 0<\nu<1,\ \ y\rightarrow\infty\}.
\label{sep-1}
\ee
It is obvious that similar statements about an almost separation of the variables are true/hold in all asymptotic domains
\be
\Omega_j^+: \{x_j\leq y_j^\nu,\ \ 0<\nu<1,\ \ y_j\rightarrow\infty\}
\label{sep-j}
\ee
in the vicinities of the screens  $\sigma_j$ by  $y_j >> x_j$.

Coming back to the domain $\Omega_1^+$, we note that the equation with an approximation potential $V_{sep}$ allows a separation of variables
$$
\left[-\Delta_{\bz} + v_1(x) + v^{eff}(y)\right]\Psi^{sep} = E\Psi^{sep}.
$$

As we are interested in finite solutions, for $\Psi^{sep}$ there naturally arises the following presentation
\be
\Psi^{sep}(\bz,\bq)=\int_{\bR^3}d\bk'\int_{\bR^3}d\bp'\psi_c(\bx,\bk')\psi_c^{eff}(\by,\bp')\delta({k'}^2+{p'}^2-E)R(\bq,\bq')+
\label{sep-chi}
\ee
$$
+\mathop{\sum}\limits_{n=1}^\infty\mathop{\sum}\limits_{l=0}^{n-1}\mathop{\sum}\limits_{m=-l}^l
\int_{\bR^3}d\bp'\psi_{nlm}(x)Y_l^m(\hx)\psi_c^{eff}(\by,\bp')\delta({p'}^2-\frac{|\alpha_1|^2}{4n^2}-E)
R_{nlm}(\bq,\bp').
$$
We use here the notations
$$
\bq=(\bk,\bp)^t,\ \ \bq'=(\bk',\bp')^t,\ \ \ q^2=E.
$$
At that $\psi_c(\bx,\bk)$ is a two-body Schrodinger operator continuous specter eigenfunction with the potential $v_1$, $\psi_{nlm}(x)Y_l^m(\hx)$ is a discrete specter eigenfunction of the same operator (by assumption the Coulomb potential $v_1$ is attractive),  $Y_l^m$ is a certain spherical function,
 $\psi_c^{eff}(\by,\bp)$  is a continuous spectrum eigenfunctions corresponding to the two-body Schrodinger operator  with the potential $v^{eff}$  (\ref{eff-1})
$$
\left(-\Delta_\bx + v_1(x)\right)\psi_c(\bx,\bk)=k^2\psi_c(\bx,\bk),
$$
$$
\left(-\Delta_\by + v^{eff}(y)\right)\psi_c^{eff}(\by,\bp)=p^2\psi_c^{eff}(\by,\bp).
$$

We start with choosing asymptotic domains in the configuration space, in which simultaneously true is a presentation for the Schrodinger operator continuous specter eigenfunctions asymptotics written in terms of variables separation and functioning in the vicinity of the screen with the index $1$, as well as the $BBK$-approximation. Introduce the domain
\be
\Omega_{\mu\nu}=\{y^{\mu}<x<y^{\nu},\ \ \frac12<\mu<\nu<1,\ \ \ y\rightarrow\infty\},
\label{mu-nu}
\ee
which arises as an intersection of the domains $\Omega_1$ (\ref{e_mu_j})  and $\Omega_1^+$  (\ref{sep-1}).
In the domain the both approximations $\Psi^{sep}$ and $\Psi^{BBK}$ for the Schrodinger operator continuous specter eigenfunction for a three-body system turns simultaneously true.

In the domain we are going to correlate two presentations described above. At that as will be shown below, an uncertainty arising by the variables separation is fixed. The uncertainty is in the structure of the kernels $R(\bq,\bq')$, $R_{nlm}(\bq,\bp')$ as well as in the structure of the spherical functions $Y_l^m$.

Note here that in the decomposition (\ref{sep-chi}) we neglect a formal decomposition in the two-body Schrodinger operator discrete specter eigenfunctions with the potential $v^{eff}(у)$ in consequence of the characteristics of the correlation $\Omega_{\mu\nu}$ parabolic domain, i.e. the coordinates $\bx$ and $\by$ disparity in the domain (the variable $\by$ can be made infinitely big on retention of the condition $x\ll y$). In fact we write in each parabolic domain a spectral decomposition only in the asymptotic channels which are actualized in it.

\section{ An "Uncertainty" Fixation and a Generating Integral}

In order to fix the uncertainty in the presentation of the spectral decomposition $\Psi^{sep}(\bz,\bq)$
(\ref{sep-chi}), we will need a certain construction to be called a generating integral.

The essence of the construction is in our presenting a certain function $g_n(\bx,\homega)$, for which there is a function $R_n(\bq,\bp',\homega)$, the one with which the following equation is true
\be
 \int_{\bS^2}d\homega g_n(\bx,\homega)R_n(\bq,\bp',\homega)=
\mathop{\sum}\limits_{l=0}^{n-1}\mathop{\sum}\limits_{m=-l}^l
\psi_{nlm}(x)Y_l^m(\hx)R_{nlm}(\bq,\bp').
\label{con-int}
\ee
The function $g_n(\bx,\homega)$ is described in detail in Appendix A. Substitute the function defined in the equation (\ref{gn-def}) into the equation (\ref{con-int}). It is easy to see that if we define the kernel $R_n(\bq,\bp',\homega)$, in its terms we will define the kernels $R_{nlm}(\bq,\bp')$ having the meaning of decomposition coefficients of the Schrodinger operator continuous specter three-body eigenfunction in discrete specter eigenfunctions of a pairwise subsystem in the presentation of the spectral decomposition (\ref{sep-chi}) type. To be exact, if we define a radial part of two-body Coulomb operator $\psi_{nlm}(x)$ eigenfunctions in the standard way \cite{Landau-3}
$$
\psi_{nlm}(x)=N_{nlm}e^{-\frac{|\alpha|}{2n}x}x^l\Phi\left(-n+l+1,2l+2,\frac{|\alpha|}{n}x\right),
$$
where $N_{nlm}$ is a normalization constant depending only on quantum numbers, then according to
(\ref{con-int}) и (\ref{f-2})-(\ref{gn-def}) we obtain the following expression for the kernels
$R_{nlm}$:
\be
R_{nlm}(\bq,\bp')=\frac{1}{N_{nlm}}\beta_{nl}
\frac{1}{(2l+1)!}
(-\frac{|\alpha|}{n})^l
 \int_{\bS^2}d\homega R_n(\bq,\bp',\homega)Y_l^m(\homega).
 \label{r-nlm}
\ee

Having thus defined the "generating" function $g_n(\bx,\homega)$, we replace a triple infinite sum in three quantum numbers by the single sum and the unit sphere integral in the spectral decomposition. Such a simplification turns quite significant, as it allows to change a search of a set of $n^2$ unknown coefficients $R_{nlm}(\bq,\bp')$ by each fixed principal quantum number $n$ for a search of one unknown function $R_n(\bq,\bp',\homega)$.
Hereafter we will be using for the generating function $g_n(\bx,\homega)$ the notation 
$\psi_n^d(\bx,\hk)$ to demonstrate that the function contains information on a set of discrete spectrum basis functions fitting the principal quantum number $n$. We will also be using a connection of the Laguerre polynomial and the confluent hypergeometric function with a whole negative first argument formed as follows
\be
\psi_n^d(\bx,\hk)=g_n(\bx,\hk)=e^{-\frac{|\alpha_1|}{2n}x}
L_{n-1}\left(\frac{|\alpha_1|}{n}x\sin^2\frac{\theta}{2}\right),\ \ \ \
\sin^2\theta/2=\frac{1-\langle\hk,\hx\rangle}{2}.
\label{psin}
\ee
Here  $L_n(y)$ -- are the Laguerre polynomials.

Thus a presentation for the function $\Psi^{sep}(\bz,\bq)$
(\ref{sep-chi}) in new terms takes form
\be
\Psi^{sep}(\bz,\bq)=\int_{\bR^3}d\bk'\int_{\bR^3}d\bp'\psi_c(\bx,\bk')\psi_c^{eff}(\by,\bp')\delta({k'}^2+{p'}^2-E)R(\bq,\bq')+
\label{sep-chi-1}
\ee
$$
+\mathop{\sum}\limits_{n=1}^\infty\int_{\bS^2}d\hk'
\int_{\bR^3}d\bp'\psi_n^d(\bx,\hk')\psi_c^{eff}(\by,\bp')\delta({p'}^2-\frac{\alpha_1^2}{4n^2}-E)
R_{n}(\bq,\bp',\hk').
$$
Now we are ready to pass on to the procedure of reconstruction of the kernels $R_{n}(\bq,\bp',\hk')$. We will be interested only in big values of the principal quantum number $n\ge M\gg 1$, i.e. in the vicinity of the Schrodinger operator discrete Coulomb specter accumulation point fitting/relating to the subsystem with index 1. Here  $M$ is a certain big whole number.

\section{Kernels $R_{n}(\bq,\bp',\hk')$ Reconstruction at $n\ge M\gg 1$}

For the reconstruction of the kernels $R_{n}(\bq,\bp',\hk')$ we will use the procedure of correlation of the approximations
$\Psi^{BBK}$ (\ref{bbk0}) and $\Psi^{sep}$ (\ref{sep-chi-1}) for a three-body operator continuous specter eigenfunction in the domain
$\Omega_{\mu\nu}$ (\ref{mu-nu}), where both approximations are true.

We will need an assumption on {$"$} nonsingular behavior \mbox{$"$}, integrability of three-body operator $\Psi_c(\bz,\bq)$ continuous specter eigenfunction in the domain
$$
x < \infty,\ \ y\gg 1.
$$
Consider a presentation for $\Psi_c(\bz,\bq)$ in the form of a spectral decomposition in the asymptotic domain
 $\Omega^+_1$ (\ref{sep-1})
\be
\Psi_c(\bz,\bq)\sim
\int_{\bR^3}d\bk'\int_{\bR^3}d\bp'\psi_c(\bx,\bk')\psi_c^{eff}(\by,\bp')
\delta({k'}^2+{p'}^2-E)R(\bq,\bq')+
\label{corr-1}
\ee
$$
+\mathop{\sum}\limits_{n'=1}^\infty\int_{\bS^2}d\hk'
\int_{\bR^3}d\bp'\psi_{n'}^d(\bx,\hk')\psi_c^{eff}(\by,\bp')\delta({p'}^2-\frac{\alpha_1^2}{4{n'}^2}-E)
R_{n'}(\bq,\bp',\hk'),\ \ \ \ \ (\bx,\by)\in\Omega^+_1.
$$
Multiply the left-hand side and the right-hand side of the equality (\ref{corr-1}) by ${\psi_n^d}^*(\bx,\hk'')$
and integrate over the variable $\bx$ в $\bR^3$. We will use the orthogonality of two-body operator $h=-\Delta_{\bx}+\frac{\alpha_1}{x}$ eigenfunctions corresponding to different spectral points.

Divide the variation domain of radial variable $x,\ 0\le x<\infty$ into three parts
$$
D_I:\ \ \ 0\le x \le y^\mu,\ \ \frac12 < \mu < 1,\ \ y\gg 1,
$$
$$
D_{II}: \ \ \ y^\mu\le x\le y^\nu,\ \ \mu<\nu<1,\ \ y\gg 1,
$$
$$
D_{III}:\ \ \ y^\nu\le x< \infty,\ \ y\gg 1.
$$
Note that all the functions $\psi_n^d(\bx,\hk)$ (\ref{psin}) exponentially decrease at infinity for any fixed $n$.
At the same time the power of exponent (in other words, the damping speed) depends on the main quantum number as
$\frac{1}{n},\ \ n\ge M\gg 1$, i.e. for big values of the main quantum number the damping takes place only at asymptotically big values $x$. From here it follows that the domain $D_{III}$ does not make a contribution into the equation (\ref{corr-1}) integration with the functions ${\psi_n^d}^*(\bx,\hk'')$.  Define the relation of the extentions of the domains $D_{II}$ и $D_{I}$:
$$
\frac{V_{D_{II}}}{V_{D_{I}}}=\frac{y^\nu-y^\mu}{y^\mu}=y^{\nu-\mu}
\left(1-O\left(\frac{1}{y^{\nu-\mu}}\right)\right)\mathop{\rightarrow}\limits_{y\rightarrow\infty}
\infty.
$$
Together with the assumption on the nonsingular character of the function $\Psi_c(\bX,\bP)$ in all the domain of integration the considerations above lead to the conclusion that the main and defining contribution into the discussed scalar product
\be
\langle\Psi_c,\psi_n^d\rangle|_{\bR^3_\bx}
\label{scal-0}
\ee
is made exactly by the domain $D_{II}$. Note that
$$
\Omega^+_1|_{D_{II}}=\Omega_{\mu\nu}.
$$
In its turn, on the domain $\Omega_{\mu\nu}$ the function $\Psi_c$ is well described by the BBK-approximation.
The last consideration allows us to redefine the scalar product (\ref{scal-0})
\be
\langle\Psi_c,\psi_n^d\rangle|_{\bR^3_\bx}\sim \langle\tilde{\Psi}^{BBK},\psi_n^d\rangle|_{\bR^3_\bx}.
\label{scal-1}
\ee
here the notation $\tilde{\Psi}^{BBK}$ is introduced for the function ${\Psi}^{BBK}$  extended in a regular way into the variable $\bx$ bound values domain.
It means that the equation (\ref{corr-1}) leads to the expression
\be
\langle\tilde{\Psi}^{BBK},\psi_n^d\rangle|_{\bR^3_\bx}(\by,\bq,\hk'')
\sim \frac{1}{2\sqrt{E+\frac{\alpha_1^2}{4{n}^2}}}
\int_{\bS^2}d\hk'
\int_{\bS^2}d\hp'_n\Sigma_n(\hk',\hk'')
\psi_c^{eff}(\by,\bp'_n)
R_{n}(\bq,\bp'_n,\hk'),
\label{scal-2}
\ee
$$
 p'_n=\sqrt{E+\frac{\alpha_1^2}{4{n}^2}},\ \ \ \ n\ge M\gg 1.
$$
We used here a notation
\be
\Sigma_n(\hk',\hk'')\equiv \int_{\bR^3}d\bx\psi_{n}^d(\bx,\hk'){\psi_{n}^d}^*(\bx,\hk'')
\label{def-pi}
\ee

The obtained equation (\ref{scal-2}) represents itself an integral equation for the kernel $R_n$. For the solution of the equation we need first to define the object $\Sigma_n$ to be called hereafter a normalizing integral.

\subsection{A Normalizing Integral }

Thus we arrive at a calculation of the normalizing integral in accordance with the expressions (\ref{def-pi}) и (\ref{psin})
\be
\Sigma_m(\hk',\hk'')=\int_{\bR^3}d\bx\psi_m^d(\bx,\hk'){\psi_m^d}^*(\bx,\hk'')
=
\label{normi}
\ee
$$
=\int_{\bS^2}d\hx\int_0^\infty dx x^2 e^{-\frac{|\alpha_1|}{m}x}
L_{m-1}\left(\frac{|\alpha_1|}{m}x\sin^2\frac{\tilde{\theta}}{2} \right)
L_{m-1}\left(\frac{|\alpha_1|}{m}x\sin^2\frac{\theta}{2} \right),\ \ \ \ m\gg 1.
$$
Here we used the notations $\cos\tilde{\theta}=\langle\hx,\hk''\rangle,\ \ \ $
$\cos{\theta}=\langle\hx,\hk'\rangle$.

After the change of the variable $t=\frac{|\alpha_1|}{m}x$ the equation (\ref{normi}) is reduced to the following form
\be
\Sigma_m(\hk',\hk'')=\frac{m^3}{|\alpha_1|^3}
\int_{\bS^2}d\hx\int_0^\infty dt t^2 e^{-t}
L_{m-1}\left(t\sin^2\frac{\tilde{\theta}}{2} \right)
L_{m-1}\left(t\sin^2\frac{\theta}{2} \right),\ \ \ \ m\gg 1.
\label{normi-1}
\ee
For a calculation of the integral on the semi-axis in the equation (\ref{normi-1}) we will use the asymptotics of the Laguerre polynomial
$L_n(x)$ by the sign $n$ \cite{Tricomi}. In dependence of the argument $x$ value we distinguish four regimes of behavior of such an asymptotics: the vicinity of zero, the oscillations domain,the transition point vicinity
$$
\upsilon=4n+2
$$
and the monotony domain.
Consider the contribution of the oscillations domain into the integral (\ref{normi-1}).

\subsubsection{The Regimes of the Asymptotics Behavior by the Laguerre Polynomials Index. The Domain of Oscillations.}

In accordance with the work \cite{Tricomi} in the oscillations domain $0<x<\upsilon=4n+2$introduce the following notations:
\be
x=\upsilon\cos^2\theta_*,\ \ \ \ 0<\theta_* < \frac{\pi}{2},
\label{tricomi-1}
\ee
$$
4\Theta=\upsilon(2\theta_*-\sin 2\theta_*)+\pi.
$$
By the fixed  $\theta_*$ Tricomi proved that the following decomposition is true
\be
e^{-\frac{x}{2}}L_n(x)=2(-1)^n(\pi\upsilon\sin 2\theta_*)^{-1/2}\times
\label{tricomi-2}
\ee
$$
\times\left[\mathop{\sum}\limits_{m=0}^{K-1}A_m(\theta_*)\left(\frac{\upsilon}{4}\sin 2\theta_*
\right)^{-m}\sin\left(\Theta+\frac{3m\pi}{2}\right)+O(n^{-K})\right],
$$
where
$$
A_0(\theta_*)=1,\ \ \ \
 A_1(\theta_*)=\frac{1}{12}\left[\frac{5}{4\sin^2\theta_*}-\sin^2\theta_*-1\right].
$$
In terms of the notations (\ref{tricomi-1})
$$
\sin 2\theta_*=2\sin \theta_*\cos \theta_*=2\left(\frac{x}{\upsilon}\right)^{1/2}
\sqrt{1-\frac{x}{\upsilon}}.
$$
At that the main term of the asymptotic series  (\ref{tricomi-2}) takes form
\be
e^{-\frac{x}{2}}L_n(x)=2(-1)^n(2\pi\upsilon)^{-1/2}\left(\frac{x}{\upsilon}\right)^{-1/4}
\left(1-\frac{x}{\upsilon}\right)^{-1/4}\times
\label{tricomi-def}
\ee
$$
\times\left(\sin\left(\frac{\upsilon\theta_*}{2}-
\frac{\upsilon}{4}\sin 2\theta_* +\frac{\pi}{4}\right)+O(1/\upsilon)\right).
$$

\subsubsection{The Domain of Monotony.}
Note the Laguerre polynomials behavior in the domain of monotony.
Following the notations \cite{Tricomi} in the domain of monotony $\upsilon<x<\infty$
introduce the following notations:
\be
x=\upsilon\ch^2\theta_*,\ \ \ \ 0<\theta_* ,
\label{tricomi-1}
\ee
$$
4\Theta=\upsilon(\sh 2\theta_* - 2\theta_*).
$$

In our calculations we will limit ourselves to the approximation (\ref{tricomi-def}).

Returning to the normalizing integral (\ref{normi-1}) we will put $n=m-1\gg 1$ and
consider the asymptotics of the expression of the form
$$
e^{-t}L_n\left(t\sin^2\frac{\theta}{2}\right)L_n\left(t\sin^2\frac{\tilde{\theta}}{2}\right) \sim
$$
$$
\sim \frac{1}{\pi\sqrt{\upsilon t}}\frac{e^{-t\left(1-\frac12(\sin^2\frac{\theta}{2}+ \sin^2\frac{\tilde{\theta}}{2} )\right)} }{\sqrt{\sin\frac{\theta}{2} \sin\frac{\tilde{\theta}}{2}}}
\left(1-\frac{t}{\upsilon}\sin^2\frac{\theta}{2}\right)^{-1/4}
\left(1-\frac{t}{\upsilon}\sin^2\frac{\tilde{\theta}}{2}\right)^{-1/4}\times
 $$
$$
\times\left\{\cos\left(\frac{\upsilon}{2}(\theta_*-\tilde{\theta}_*)-
\frac{\upsilon}{4}(\sin 2\theta_*-\sin 2\tilde{\theta}_*)\right)-
\cos\left(\frac{\upsilon}{2}(\theta_*+\tilde{\theta}_*)-
\frac{\upsilon}{4}(\sin 2\theta_*+\sin 2\tilde{\theta}_*)+\frac{\pi}{2}\right)\right\}.
 $$
We use here the notations
 $$
 \theta_*=\arccos\sqrt{\frac{t}{\upsilon}}\sin\frac{\theta}{2},\ \ \
\tilde{\theta}_*=\arccos\sqrt{\frac{t}{\upsilon}}\sin\frac{\tilde{\theta}}{2}.
 $$
 Thus the normalizing integral (\ref{normi-1}) takes form
 \be
\Sigma_n(\hk',\hk'')\sim
\frac{2^6 n^3}{\pi\sqrt{\upsilon}|\alpha_1|^3}\int_{\Delta}^{\upsilon-\Delta} dt t^{3/2}
\int_{\bS^2}d\hx
e^{-t\left(1-\frac12(\sin^2\frac{\theta}{2}+ \sin^2\frac{\tilde{\theta}}{2} )\right)}\times
\label{int-10}
\ee
$$
\times\frac{1}{\sqrt{\sin\frac{\theta}{2} \sin\frac{\tilde{\theta}}{2}}}
\left(1-\frac{t}{\upsilon}\sin^2\frac{\theta}{2}\right)^{-1/4}
\left(1-\frac{t}{\upsilon}\sin^2\frac{\tilde{\theta}}{2}\right)^{-1/4}\times
$$
$$
\times
\left\{\cos\left(\frac{\upsilon}{4}\left(2(\theta_*-\tilde{\theta}_*)-
(\sin 2\theta_*-\sin 2\tilde{\theta}_*)\right)\right)+
\sin\left(\frac{\upsilon}{4}\left(2(\theta_*+\tilde{\theta}_*)-
(\sin 2\theta_*+\sin 2\tilde{\theta}_*)\right)\right)\right\}.
$$
Here the parameter $\Delta$ within limits of integration is connected to defining the bounds of the oscillations domain:
$\Delta=O(\upsilon^\varrho),\ \ 0<\varrho<\frac13$.
In the Cartesian coordinate system $(x,y,z)$ the unit vector $\hk''$ is directed along the axis $z$,
the unit vector $\hx$ is characterized by the pair of angles $(\theta,\ \varphi)$ in the corresponding spherical coordinate system. The unit vector $\hk'$ is characterized by the pair of angles $(\theta_{k'},\ \varphi_{k'})$.
At that
\be
\cos\tilde{\theta}=\sin\theta\sin\theta_{k'}\cos(\varphi-\varphi_{k'})
+\cos\theta\cos\theta_{k'},
\label{connect}
\ee
$$
0\le\theta\le\pi,\ \ \ 0\le\theta_{k'}\le\pi,\ \ \ 0\le\varphi\le 2\pi,\ \ \ 0\le\varphi_{k'}\le 2\pi,
$$

Consider the angular variables integral(the $d\hx$ integral) as an integral in dependence on the extraneous variable $t$ as on the parameter.
Consider the sinus argument, which is the multiplier by the big parameter $\frac{\upsilon}{4}$ or, to put it in other way, a phase function
\be
S(\theta,\varphi)=2\arccos\left(\sqrt{\frac{t}{\upsilon}}\sin\frac{\theta}{2}\right)+
2\arccos\left(\sqrt{\frac{t}{\upsilon}}\sin\frac{\tilde{\theta}}{2}\right)-
\label{phase-0}
\ee
$$
-2\sqrt{\frac{t}{\upsilon}}\sqrt{1-\frac{t}{\upsilon}\sin^2\frac{\theta}{2}}\sin\frac{\theta}{2}-
2\sqrt{\frac{t}{\upsilon}}\sqrt{1-\frac{t}{\upsilon}\sin^2\frac{\tilde{\theta}}{2}}\sin\frac{\tilde{\theta}}{2}.
$$
Note that by the small arguments ($0<\frac{t}{\upsilon}< 1$) the function $\arccos$ behaves as follows
$$
\arccos\left(\sqrt{\frac{t}{\upsilon}}f\right)=\frac{\pi}{2}-\sqrt{\frac{t}{\upsilon}}f+
O\left(\left(\frac{t}{\upsilon}\right)^{3/2}f^{3/2}\right).
$$
Therefore the dependent on angles function in the argument of sinus and cosinus is multiplied on large parameter 
$\sqrt{\upsilon t}$, where
\be
\Delta< t<\upsilon-\Delta,\ \ \Delta=O(\upsilon^\varrho).
\label{lim-cond}
\ee
At the same time the angle-dependent function in the decreasing exponent is multiplied by the parameter $t$. It is obvious that for the big parameters in the power of the oscillating and decreasing exponents as coefficients in expressions in dependence on angular variables to be equal, we must require
\be
t=O(\upsilon),\ \ \ \upsilon\rightarrow\infty.
\label{upsi}
\ee
The term, however, as it follows from the angular function structure in the power of the decreasing exponent
in the expression (\ref{int-10}) - a non-negative and vanishing only in the point on the plane $(\theta,\varphi)$
defined by the condition 
\be
\theta=\tilde{\theta}=\pi
\label{cond-2}
\ee
can be realized only within an arbitrarily small vicinity of the point defined by the condition (\ref{cond-2}).

In its turn, the condition (\ref{cond-2}) in the point can be realized only when fulfilling a supplementary term already on the external parameters
$$
\hk'=\hk''.
$$
In the contrary case, the variable $t$ outer integral convergence in the expression (\ref{int-10})
will occur well before the condition (\ref{upsi}) starts being fulfilled in consequence of the expression under integral sign exponentially decrease.

Note too that in the initial presentation of the normalizing integral (\ref{normi-1})
the point (\ref{cond-2}) in the angular subspace a priori is not singular for any value of the index $m$.
Thus we will separate an arbitrarily small vicinity of the point defined by the condition (\ref{cond-2}) and will regard the expression
(\ref{int-10}) as an integral by the exterior of the vicinity wherever required.

We have come to the conclusion that the big parameter in the angular variables integral in the expression (\ref{int-10}) can be found/is contained only in fast oscillating functions. From these considerations calculate this integral in accordance with Appendix B:
\be
\Sigma_n(\hk',\hk'')\sim
\Im\left\{\frac{2^4 n^4}{|\alpha_1|^3}
\int_{0}^{1} ds s
e^{-4ns\sin^2\frac{\theta_{k'}}{4}}
\frac{e^{inS(\theta_0,\varphi_0)+i\frac{\pi}{2}}}
{\cos\frac{\theta_{k'}}{4}\sqrt{\cos\frac{\theta_{k'}}{2}}
\sqrt{1-s\cos^2\frac{\theta_{k'}}{4}}\sqrt{1-s\cos\frac{\theta_{k'}}{2}}}
\right\}.
\label{int-13}
\ee
Returning to the equation(\ref{scal-2}) we can now on account of the arisen simplifications
pass on to the next step and calculate the unit sphere $d\hk'$ integral:
\be
T_n(\hk'',\bp'_n,\bq)\equiv
\int_{\bS^2}d\hk'
\Sigma_n(\hk',\hk'')
R_{n}(\bq,\bp'_n,\hk').
\label{scal-3}
\ee
To do it, pass on to the variable $\alpha$, $\ \ \cos\alpha=\langle\hk',\hk''\rangle$:
\be
 T_n(\hk'',\bp'_n,\bq)=\frac{2^4 n^4}{|\alpha_1|^3}
 \int_0^{2\pi}d\varphi \int_{0}^\pi
 \frac{d\alpha\sin\alpha}{\cos\frac{\alpha}{4}\sqrt{\cos\frac{\alpha}{2}}}
\int_{0}^{1}s ds
\frac{\cos(nS(\theta_0,\varphi_0))e^{-4ns\sin^2\frac{\alpha}{4}}}
{\sqrt{1-s\cos^2\frac{\alpha}{4}}\sqrt{1-s\cos\frac{\alpha}{2}}}
R_{n}(\bq,\bp'_n,\hk'),\ \ \ \ \hk'=(\alpha,\varphi),
\label{int-k}
\ee
where 
$$
S(\theta_0,\varphi_0)=4\arccos\left(\sqrt{s}\cos\frac{\alpha}{4}\right)-4\sqrt{s}
\sqrt{1-s\cos^2\frac{\alpha}{4}}\cos\frac{\alpha}{4}.
$$

Note that we cannot consider the expression (\ref{int-k}) as an iterated integral,
as integrating, for example, of variable $\alpha$, and separating of the stationary point $\alpha=0$ contribution
leads to an occurrence of a variable $s$ non-integrable pole singularity, $\ s=1$.
Thus we should regard the expression (\ref{int-k}) as a dual integral separating explicitly the vicinity
of the point $(0,1)$ on the plane $(\alpha,s)$. It is the vicinity of this point, as will be demonstrated below,
that makes a main contribution into the integral.

\subsection{The Calculation of the Normalization Integral.}

The main idea of the next step is in introducing new variables on the plane $(\alpha,s)$
in the most convenient way, so that the contribution of the point vicinity into the integral is described.
By changing
\be
s=1-\frac{\zeta^2}{16},\ \ \ \ \ 0\le\zeta\le 4
\label{uv-s}
\ee
we  will transfer into the origin of coordinates $(0,0)$ on the plane $(\alpha,\zeta)$
the point $(0,1)$ on the plane $(\alpha,s)$ and introduce a pole system of coordinates in the unit
vicinity of the point $(0,0)$
\be
\alpha=\rho\sin\omega,\ \ \ \ \ \zeta=\rho\cos\omega,
\label{polar}
\ee
$$
0\le\rho\le 1,\ \ \ \ 0\le\omega\le\frac{\pi}{2}.
$$
It is easy to demonstrate that the exterior integral of the specified point $(0,0)$ vicinity will make
a contribution of the following order of vanishing into the expression (\ref{int-k}). Note as well that by $\rho\ll 1$
the expression (\ref{int-k}) is formed as
\be
 T^{\delta}_n(\hk'',\bp'_n,\bq)=\frac{2^4 n^4}{|\alpha_1|^3}
 \int_0^{2\pi}d\varphi\int_0^{\delta}\rho d\rho \int_{0}^{\pi/2}d\omega\sin\omega\cos\omega
 \left(1-\frac{\rho^2}{16}\cos^2\omega\right)\times
 \label{delta-10}
\ee
$$
\times\frac{\cos(n\rho^3A(\omega))}{\sqrt{1+\sin^2\omega}}
\exp\left\{-n\rho^2\sin^2\omega(1-\frac{\rho^2}{16}\cos^2\omega)\right\}
\tilde{R}_{n}(\bq,\bp'_n,\rho\sin\omega,\varphi),
$$
where $A(\omega)=\frac{1}{32}(1-\frac12\sin^2\omega\cos^2\omega)$. The two last arguments of the function
$\tilde{R}_{n}$
correspond to the angular variables of the vector $\hk'$
in the spherical coordinate system, in which zero of the first angular variable corresponds to the collinearity of the vectors $\hk'$ and $\hk''$. The sign $\ \tilde{}\ $ in the notation $\tilde{R}_{n}$ is connected with a transition to angular variables characterizing the third vector variable function $R_n$.

We will not make the specified in (\ref{polar}) change of variables explicitly but will perform the specified calculations in a general way, having reformulated the expression  (\ref{int-k}) in the following form:
\be
 T_n(\hk'',\bp'_n,\bq)=\frac{2^4 n^4}{|\alpha_1|^3} \int_0^{2\pi}d\varphi
\int_0^1 du \int_0^1 dv e^{-4n\beta uv f(u,v)}\cos(4nu^{3/2}g(u,v))F(u,v).
\label{uv-1}
\ee
Here 
\be
u=\rho^2,\ \ \ \ \ v=\sin^2\omega,\ \ \ \ \beta=\frac{1}{16}.
\label{uv-00}
\ee
Also the notations are used
$$
f(u,v)=(1+\phi(u,v))(1-\beta u(1-v)),\ \ \ \phi(u,v)\mathop{=}\limits_{u,v\rightarrow 0}O(uv),
$$
$$
g(u,v)=\frac{13}{6}+v(1-v)+ \sigma(u,v),\ \ \ \sigma(u,v)\mathop{=}\limits_{u\rightarrow 0}O(u^{1/2}).
$$
The functions  $F(u,v)$,  $f(u,v)$ and  $g(u,v)$ arise as a result of two consecutive changes of the variables
(\ref{uv-s})-(\ref{polar}) and (\ref{uv-00}) in the expression (\ref{int-k}).
The functions $F(u,v)$,  $\phi(u,v)$ and  $\sigma(u,v)$ are bounded, twice differentiable functions of their arguments.
We assume here that the function $F(u,v)$ contains also a kernel $\tilde{R}_n$,
\be
F(u,v)=\hat{F}(u,v)\tilde{R}_{n}(\bq,\bp'_n,\sqrt{uv},\varphi),
\label{r-uv}
\ee
which in its turn possesses bounded derivatives on all variables of integration.
At that the expression $\hat{F}(u,v)$ arises actually in connection with the change of
variables in the kernel of the integral (\ref{int-k}). As already discussed above, two last arguments of the function
$\tilde{R}_{n}(\bq,\bp'_n,\sqrt{uv},\varphi)$ correspond to the vector $\hk'$ angular variables in the spherical coordinate system, where zero of the first angular variable relates to the collinearity of the vectors $\hk'$ and $\hk''$.

In Appendix C the structure of the expression $T_n(\hk'',\bp'_n,\bq)$ is described. Redefine the coefficients 
$T_n(\hk'',\bp'_n,\bq)$
described in Appendix C in accordance with the factorization (\ref{r-uv}):
\be
D_i(G)=\hat{D}_i(G)R_{n}(\bq,\bp'_n,\hk''),\ \ \ \ i=1,2.
\label{fact-uv}
\ee
Note that the factorization (\ref{fact-uv}) turns possible, as according to (\ref{r-uv})
the expression  $F(0,v)$, (the one describing the asymptotics main order in accordance with Appendix C) contains the multiplier $\tilde{R}_{n}(\bq,\bp'_n,0,\varphi)$ independent on the integration variables $u$ and $v$.

Introduce also the coefficients $B_i,\ \ i=1,2$ in the following way
\be
B_i=\frac12(\hat{D}_i(G)+\hat{D}_i(G^*)),\ \ \ \ i=1,2.
\label{def-b12}
\ee

According to Appendix C
and (\ref{uv-1}),(\ref{scal-3})
\be
 T_n(\hk'',\bp'_n,\bq)=\int_{\bS^2}d\hk'
\Sigma_n(\hk',\hk'')
R_{n}(\bq,\bp'_n,\hk')=
n^3(B_2\ln n + B_1  + o(1) )\frac{2^5\pi}{|\alpha_1|^3} R_{n}(\bq,\bp'_n,\hk'').
 \label{res-1}
 \ee
It means that the equation (\ref{scal-2}) takes form 
\be
\langle\tilde{\Psi}^{BBK},\psi_n^d\rangle|_{\bR^3_\bx}(\by,\bq,\hk'')
\sim n^3\frac{\left( B_2\ln n + B_1 \right)}{2\sqrt{E+\frac{\alpha_1^2}{4{n}^2}}}
\frac{2^5\pi}{|\alpha_1|^3}
\int_{\bS^2}d\hp'_n
\psi_c^{eff}(\by,\bp'_n)
R_{n}(\bq,\bp'_n,\hk'').
\label{scal-20}
\ee
$$
 p'_n=\sqrt{E+\frac{\alpha_1^2}{4{n}^2}},\ \ \ \ n\ge M\gg 1.
$$

The equation  (\ref{res-1}) demonstrates an effectively singular at angular variable 
$\hk'$ structure of the normalization integral (\ref{normi}). In other words, we can say that
\be
\Sigma_n(\hk',\hk'')\sim \delta(\hk'',\hk').
\label{s-delta}
\ee
This result at first sight is remarkable as in the expression (\ref{normi}) we normalize discrete spectrum functions. The explanation of the fact is connected with the condition that we normalize in the expression (\ref{normi}) functions from the discrete spectrum accumulation point spectral vicinity of the two-body Coulomb Schroedinger operator ($n\gg 1$).

Consecutively, in a spectral sense these functions are closer to the continuous spectrum functions which are normalized exactly on the $\delta$-function and are angular variable generalized functions.

Consider now the behaviour of the left-hand side of the equation (\ref{scal-20})

\subsection{ The Behaviour of the Absolute Term of the Equation for the Kernel $R_n$ }

We study here the asymptotics of the scalar product
$\langle\tilde{\Psi}^{BBK},\psi_n^d\rangle|_{\bR^3_\bx}(\by,\bq,\hk'')$ by large values of the variable $y\gg 1$ and by 
large values of the index  $n\gg 1$.

The structure of the asymptotic expression is described in Appendix D and is formed as follows:
\be
\langle\tilde{\Psi}^{BBK},\psi_n^d\rangle|_{\bR^3_\bx}(\by,\bq,\hk'')\sim
iB_0^{in}(\bq)\frac{2\pi}{iyp}\delta(\hy,-\hp)\frac{1}{y}e^{-iyp+i\omega\ln y}
Z^{in}(\bq,\hk'')-
\label{q-def-100}
\ee
$$
-iB_0^{out}(\bq)\frac{2\pi}{iyp}\delta(\hy,\hp)\frac{1}{y}e^{iyp+i\omega\ln y}
Z^{out}(\bq,\hk'').
$$
Here 
$$
Z^{in(out)}(\bq,\hk'')=-\frac{N_c^{(1)}\Gamma(3+i\eta)}{k^{4+i\eta}}\left[
e^{\frac{\pi\eta}{2}}\langle\hk,\bL_{in(out)}(\bq)\rangle
\Phi\left(3+i\eta,1,i\frac{|\alpha_1|}{2k}(1+\langle\hk'',\hk\rangle)\right)
-e^{-\frac{\pi\eta}{2}}H^{in(out)}(\bq,\hk'')\right],
$$
where 
$$
H^{in(out)}(\bq,\hk'')\equiv \int_{\bS^2}d\hx\langle\hx,\bL_{in,(out)}(\bq)\rangle
\Phi\left(3+i\eta,1,-i\frac{|\alpha_1|}{2k}(1-\langle\hk'',\hx\rangle)\right)
s_c(\hx,\bk).
$$

Thus the expression (\ref{q-def-100})described the behavior of the left-hand side of the equation
(\ref{scal-20}) for the kernel $R_n(\bq,\bp'_n,\hk'')$.

\subsection{ Ansatz for the Kernel $R_n(\bq,\bp'_n,\hk')$ and the Solution of the Equations for the Kernels $r_n^{in,out}(\bq,\bp'_n)$. }

In accordance with the structure of the expression (\ref{q-def-100}) we will search for the structure of the kernel $R_n(\bq,\bp'_n,\hk')$
in the form of the sum of two terms correlated correspondingly to the converging and diverging waves.
We will search for the kernel $R_n(\bq,\bp'_n,\hk')$ in the following form
\be
R_n(\bq,\bp'_n,\hk')=r_n^{in}(\bq,\bp'_n)Z^{in}(\bq,\hk'')+r_n^{out}(\bq,\bp'_n)Z^{in}(\bq,\hk').
\label{kern-y}
\ee
Rewrite the expression (\ref{scal-20}) in the form of the system of two equations for the kernels
$r_n^{in}(\bq,bp'_n)$ и $r_n^{out}(\bq,\bp'_n)$
\be
w_n\int_{\bS^2}d\hp'_n
\psi_c^{eff}(\by,\bp'_n)r_n^{in}(\bq,\bp'_n)=
iB_0^{in}(\bq)\frac{2\pi}{iyp}\delta(\hy,-\hp)\frac{1}{y}e^{-iyp+i\omega\ln y},
\label{eq-y-in}
\ee
\be
w_n\int_{\bS^2}d\hp'_n
\psi_c^{eff}(\by,\bp'_n)r_n^{out}(\bq,\bp'_n)=
iB_0^{out}(\bq)\frac{2\pi}{iyp}\delta(\hy,\hp)\frac{1}{y}e^{iyp+i\omega\ln y},
\label{eq-y-out}
\ee
where the notation is introduced 
$$
w_n\equiv n^3\frac{\left( B_2\ln n + B_1 \right)}{2\sqrt{E+\frac{\alpha_1^2}{4{n}^2}}}
\frac{2^5\pi}{|\alpha_1|^3}.
 $$
Start with the solution of the equation (\ref{eq-y-in}). Introduce a Cartesian coordinate system in which the vector $\hp$ is directed along the axis $Z$. Let the direction of the axis $X$ coincides with the direction of the vector $\bk$ projection onto the plane orthogonal to the vector $\bp$. Assuming $y\gg 1$, rewrite the left-hand side of the equation (\ref{eq-y-in}) in the spherical coordinate system corresponding to the given Cartesian system:
\be
L^{in}\equiv w_n\int_0^{2\pi}d\varphi\int_{-1}^1dt
e^{iy\sqrt{E}(t\cos\theta_{yp}+\sqrt{1-t^2}\sin\theta_{yp}\cos(\varphi-\varphi_{yk}))}
e^{i\eta^{eff}\ln y}\tilde{r}_n^{in}(\bq,t,\varphi).
\label{eq-y-in-1}
\ee
Here the notations are introduced $t=\cos\theta=\langle\hp,\hp'_n\rangle$,
$\cos\theta_{yp}=\langle\hp,\hy\rangle$. The angles $\varphi$ and  $\varphi_{yk}$ are azimuthal angles corresponding to the vectors $\hp'_n$ and $\hy$ accordingly. The kernel $\tilde{r}_n^{in}(\bq,t,\varphi)$ is produced/generated by the kernel $r_n^{in}(\bq,\bp'_n)$ by the transition to the spherical coordinates of the vector $\hp'_n$.
We will search for the function $\tilde{r}_n^{in}(\bq,t,\varphi)$ in the following form
\be
\tilde{r}_n^{in}(\bq,t,\varphi)=\kappa_{0}^{(in)}\left(t-\frac{p}{\sqrt{E}}\right)_+^{ib}
\mbox{ce}_{2l}(\varphi,s).
\label{kern-mat-in}
\ee
Here $\kappa_{0}^{(in)}$ is a normalization constant, $b$ is a parameter
$b=\eta^{eff}-\omega$. The notation $\left(t-\frac{p}{\sqrt{E}}\right)_+^{ib}$ 
is introduced for the distribution $\chi_+^\lambda$ \cite{GSh}.
The function $\mbox{ce}_{2l}(\varphi,s)$ is a Mathieu function \cite{Matie}:
\be
\mbox{ce}_{2l}(\varphi,s)=\mathop{\sum}\limits_{r=0}^\infty A_{2r}^{(2l)}\cos 2r\varphi,
\label{matie-00}
\ee
The decomposition coefficients $ A_{2r}^{(2l)}$ are defined in a recurrent way according to (8.60)
\cite{Grad},  $s$ is a certain real-valued parameter.

Note that the emergence of Mathieu functions in the kernel (\ref{kern-mat-in}) structure is connected with a necessity to damp out high-frequency oscillations generated by the non-coincidence of the unit vectors $\hp'_n$ and $\hp$ directions.

The deviation in its turn is connected to the situation that the two-body basis functions $\psi^{eff}(\by,\bp'_n))$ relating to the "big" variable $\by$ correspond to the point ${\bp'_n}^2=E$ spectral vicinity. At the same time the absolute value of momentum conjugate to the variable  $\by$  in the expression of the three-body distorted wave $\Psi^{BBK}$ equals to $p$ ($p^2< E$).

As we saw above, the functions $\psi^{eff}(\by,\bp'_n))$ relating to the Coulomb discrete spectrum accumulation point vicinity make a contribution into the approximation of the three-body wave $\Psi^{BBK}$ only of the correction order  ($\frac{1}{y^2}$ in a generalized sense).

To separate the contribution, it proves to be necessary to perform azimuthal angle averaging with Mathieu functions.
Note that the Mathieu functions, according to (\ref{matie-00}), represent themselves decompositions including those by fast oscillating functions.

Substitute the kernel (\ref{kern-mat-in}) into the equation (\ref{eq-y-in-1}) and integrate by the variable $\varphi$, using the equations \cite{Grad} (6.925.1)-(6.925.2):
\be
\int_0^{2\pi}\sin[z_1\cos(\varphi-\alpha)]\mbox{ce}_{2l}(\varphi,s)d\varphi=0,
\label{matie-s}
\ee
\be
\int_0^{2\pi}\cos[z_1\cos(\varphi-\alpha)]\mbox{ce}_{2l}(\varphi,s)d\varphi=
\frac{2\pi A_0^{(2l)}}{\mbox{ce}_{2l}(0,s)\mbox{ce}_{2l}(\frac{\pi}{2},s)}
\mbox{Ce}_{2l}(\zeta,s)\mbox{ce}_{2l}(\vartheta,s).
\label{matie-c}
\ee
Here the notations were used 
$\mbox{Ce}_{2l}(\zeta,s)=\mathop{\sum}\limits_{r=0}^\infty A_{2r}^{(2l)}\ch 2r\varphi$,
\be
z_1=2\sqrt{s}\sqrt{\ch^2\zeta-\sin^2\vartheta},\ \ \ \ \ \tg\alpha=\th\zeta\tg\vartheta.
\label{fix-zm}
\ee
The function $\mbox{Ce}_{2l}(\zeta,s)$ is an associated Mathieu function of the first kind.
Introducing the notations
\be
z_1=y\sqrt{E}\sqrt{1-t^2}\sin\theta_{yp},\ \ \ \alpha=\varphi_{yk},
\label{matie-zal}
\ee
we fix the parameters  $\zeta$ and  $\vartheta$ in accordance with the equations (\ref{fix-zm}).
Substituting in the equation (\ref{eq-y-in-1}) the variable  $\sigma=t-\frac{p}{\sqrt{E}}$,
rewrite  (\ref{eq-y-in-1}) in the form of 
\be
L^{in}\sim \kappa_{0}^{(in)} y^{i\eta^{eff}}
\frac{2\pi w_n A_0^{(2l)}}{\mbox{ce}_{2l}(0,s)\mbox{ce}_{2l}(\frac{\pi}{2},s)}
  e^{iyp\cos\theta_{yp}}
\int_0^\infty d\sigma \sigma^{ib}_+
e^{iy\sqrt{E}\sigma\cos\theta_{yp}}
\mbox{Ce}_{2l}(\tilde{\zeta},s)\mbox{ce}_{2l}(\tilde{\vartheta},s).
\label{eq-y-in-11}
\ee
The variables  $\tilde{\zeta}$ and $\tilde{\vartheta}$ are generated by the original variables 
${\zeta}$ and ${\vartheta}$ when changing  $\sigma=t-\frac{p}{\sqrt{E}}$.

Integrate the left and right-hand sides of the equation (\ref{eq-y-in-11}) with an infinitely differentiable function $H(\hp)$ defined on the unit sphere
\be
\langle L^{in},H\rangle_{\bS^2_\hp}\sim \kappa_{0}^{(in)}
y^{i\eta^{eff}}
\frac{2\pi w_n A_0^{(2l)}}{\mbox{ce}_{2l}(0,s)\mbox{ce}_{2l}(\frac{\pi}{2},s)}\times
\label{eq-y-in-12}
\ee
$$
\times
\int_0^{2\pi}d\varphi_{yp}\int_0^\pi d\theta_{yp}\sin\theta_{yp}\tilde{H}(\theta_{yp},\varphi_{yp})
  e^{iyp\cos\theta_{yp}}
\int_0^\infty d\sigma \sigma^{ib}_+
e^{iy\sqrt{E}\sigma\cos\theta_{yp}}
\mbox{Ce}_{2l}(\tilde{\zeta},s)\mbox{ce}_{2l}(\tilde{\vartheta},s).
$$
Taking into account that $\Im p< 0$, find that the main contribution into the integral on $d\theta_{yp}$ is defined by the point $\theta_{yp}=\pi$.

In accordance with the equation (\ref{matie-zal}) in this case $z_1=0$. From the equation (\ref{fix-zm}) it follows that
$$
\zeta=0,\ \ \ \ \vartheta=\frac{\pi}{2},\ \ \ \ \
$$
and, as a consequence, $\alpha=\varphi_{yk}=\frac{\pi}{4}$ or $\alpha=\frac{5\pi}{4}$.
The last statement allows us to make conclusions about the geometry of the averaging performed.

Thus the equation (\ref{eq-y-in-12}) takes form
\be
\langle L^{in},H\rangle_{\bS^2_\hp}\sim \kappa_{0}^{(in)}
y^{i\eta^{eff}}
\frac{4\pi^2 w_n A_0^{(2l)}}{-ipy}H(-\hy)
  e^{-iyp}
\int_0^\infty d\sigma\sigma^{ib}_+
e^{-iy\sqrt{E}\sigma}.
\label{eq-y-in-13}
\ee
Having changed the variable $\rho=y\sigma$ 
we obtain according to (3.381.4) \cite{Grad}
\be
\langle L^{in},H\rangle_{\bS^2_\hp}\sim\kappa_{0}^{(in)}
\frac{4\pi^2 w_n A_0^{(2l)}}{-ip}\frac{\Gamma(1+ib)}{(i\sqrt{E})^{1+ib}}
\frac{e^{-iyp+i\omega\ln y}}{y^2}H(-\hy).
\label{eq-y-in-14}
\ee
Comparing the equation obtained and the equation (\ref{eq-y-in}), we find the normalization variable
$\kappa_{0}^{(in)}$
$$
\kappa_{0}^{(in)}=\frac{B_0^{in}(\bq)\sqrt{E}^{1+ib}e^{\frac{\pi b}{2}}}
{2\pi w_n A_0^{(2l)}\Gamma(1+ib)}.
$$

We come to the conclusion that in the sense of generalized functions the kernel (\ref{kern-mat-in})
satisfies the equation (\ref{eq-y-in}).

Similarly we can define the kernel $r_n^{out}(\bq,\bp'_n)$
\be
\tilde{r}_n^{out}(\bq,t,\varphi)=\kappa_{0}^{(out)}\left(t-\frac{p}{\sqrt{E}}\right)^{ib}_+
\chi_{p}(t)\mbox{ce}_{2l}(\varphi,s).
\label{kern-mat-out}
\ee
and demonstrate that it satisfies the equation (\ref{eq-y-out}).

Formulate the final result in the following form.
The kernel $R_n(\bq,\bp'_n,\hk')$  looks as follows 
\be
\tilde{R}_n(\bq,t,\varphi,\hk')=\frac{\varpi_{0}^{(in)(\bq)}}{n^3(B_2\ln n+B_1)}
\left(t-\frac{p}{\sqrt{E}}\right)^{ib}_+
\chi_{p}(t)\mbox{ce}_{2l}(\varphi,s)Z^{in}(\bq,\hk')+
\label{kern-y00}
\ee
$$
+\frac{\varpi_{0}^{(out)}(\bq)}{n^3(B_2\ln n+B_1)}\left(t-\frac{p}{\sqrt{E}}\right)^{ib}_+
\chi_{p}(t)\mbox{ce}_{2l}(\varphi,s)Z^{out}(\bq,\hk').
$$
Here the notations were used 
$$
Z^{in(out)}(\bq,\hk')=-\frac{N_c^{(1)}\Gamma(3+i\eta)}{k^{4+i\eta}}\left[
e^{\frac{\pi\eta}{2}}\langle\hk,\bL_{in(out)}(\bq)\rangle
\Phi\left(3+i\eta,1,i\frac{|\alpha_1|}{2k}(1+\langle\hk',\hk\rangle)\right)
-e^{-\frac{\pi\eta}{2}}H^{in(out)}(\bq,\hk')\right],
$$
where
$$
H^{in(out)}(\bq,\hk')\equiv \int_{\bS^2}d\hx\langle\hx,\bL_{in,(out)}(\bq)\rangle
\Phi\left(3+i\eta,1,-i\frac{|\alpha_1|}{2k}(1-\langle\hk',\hx\rangle)\right)
s_c(\hx,\bk).
$$
Here  $t=\langle\hp,\hp'_n\rangle$, the variable $\varphi$ is an angle between the vectors 
$[\hp\times\hk]$ and  $[\hp\times\hp'_n]$, measured in the positive direction upon the condition that the vector triple $[\hp\times\hk]$, $\ [\hp\times\hp'_n]$ and $\hp$ are positively oriented,
$$
\varpi_{0}^{(in,out)}(\bq)=B_0^{in,out}(\bq)\frac{|\alpha_1|^3E^{1+i\frac{b}{2}}e^{\frac{\pi b}{2}}}
{2^5\pi^2A_0^{(2l)}\Gamma(1+ib)}.
$$

\section{Conclusions and Results.}

Formulate the final conclusions. Come back to the spectral decomposition (\ref{corr-1}) in the domain $\Omega_1^+$
and separate a contribution of the pairwise Coulomb operator discrete spectrum accumulation point vicinity. At that the contribution into the continuous spectrum of the pairwise Coulomb operator does not make a contribution into the spectral decomposition. The contribution of the discrete spectrum is bounded by summing over the main quantum number in the expression (\ref{corr-1}) from a certain value $M,\  M\gg 1$ to infinity. The value of the parameter $M$ is dictated by the energy distribution between subsystems in the initial conditions (by the relation of values $k^2$ and $p^2$):
\be
\Psi_c^{acc}(\bz,\bq)\sim
\mathop{\sum}\limits_{n'=M}^\infty \frac{1}{2\sqrt{E+\frac{|\alpha_1^2|}{4{n'}^2}}}\int_{\bS^2}d\hk'
\int_{\bS^2}d\hp'_n\psi_{n'}^d(\bx,\hk')\psi_c^{eff}(\by,\bp'_n)
R_{n'}(\bq,\bp'_n,\hk'),
\label{corr-100}
\ee
$$
(\bx,\by)\in\Omega^+_1,\ \ \ \ p'_n=\sqrt{E+\frac{|\alpha_1^2|}{4{n'}^2}}.
$$
The expression for the kernel $R_{n'}(\bq,\bp'_n,\hk')$ is defined in the equation (\ref{kern-y00}).
According to the expression  (\ref{psin})
$$
\psi_n^d(\bx,\hk)=g_n(\bx,\hk)=e^{-\frac{|\alpha_1|}{2n}x}
L_{n-1}\left(\frac{|\alpha_1|}{n}x\sin^2\frac{\theta}{2}\right),\ \ \ \
\sin^2\theta/2=\frac{1-\langle\hk,\hx\rangle}{2},
$$
where $L_n(y)$ are the Laguerre polynomials.

Effectively the expression (\ref{corr-100}) is reduced to the following form 
\be
\Psi_c^{acc}(\bz,\bq)\sim \mathop{\sum}\limits_{n=M}^\infty
\frac{1}{n^3(B_2\ln n+B_1)}e^{-\frac{|\alpha_1|}{2n}x}\int_{\bS^2}d\hk'
L_{n-1}\left(\frac{|\alpha_1|}{2n}x(1-\langle\hk',\hx\rangle)\right)
U(\by,\bq,\hk'),
\label{main-1}
\ee
where $U(\by,\bq,\hk')$ is a smooth function of the variable  $\hk'$, while the parameter $x$
takes on arbitrary values on the positive semi-axis ($x\in[0,\infty$).
Introduce a new parameter $t_n=\frac{x}{n^2}$.  At that let the values of the parameter $t_n$ be such that the Laguerre polynomial argument can be big enough
$$
0<\frac{n|\alpha_1|}{2}t_n(1-\langle\hk',\hx\rangle)< 4n+2,\ \ \ \ n\ge M\gg 1.
$$
In other words, we want to consider the Laguerre polynomial asymptotics in the oscillations domain.
The asymptotics according to the above said is described by the expressions
(\ref{tricomi-1})-(\ref{tricomi-def}). According to the expression (\ref{tricomi-def})
\be
e^{-\frac{r}{2}}L_n(r)=2(-1)^n(2\pi\upsilon)^{-1/2}\left(\frac{r}{\upsilon}\right)^{-1/4}
\left(1-\frac{r}{\upsilon}\right)^{-1/4}
\left(\sin\left(\frac{\upsilon\theta_*}{2}-
\frac{\upsilon}{4}\sin 2\theta_* +\frac{\pi}{4}\right)+O(1/\upsilon)\right),
\label{tricomi-def-00}
\ee
$$
\upsilon=4n+2,\ \ \ \
r=\upsilon\cos^2\theta_*,\ \ \ \ 0<\theta_* < \frac{\pi}{2}
$$
 in the oscillations domain the Laguerre polynomials asymptotics contains an exponential growth and a fast oscillating function. At that if we find ourselves outside/beyond the stationary point vicinity (when integrating by $d\hk'$ in the equation (\ref{main-1}) ), the point
 $$
\langle\hk',\hx\rangle = -1
$$
corresponds to one of the bounds of the integration domain and, accordingly, gives one of the main contributions into the integral (\ref{main-1}). In addition the point (in the Laguerre polynomial oscillations domain) fully compensates the exponential decrease $e^{-\frac{|\alpha_1|}{2n}x}$ present in the equation (\ref{main-1}) and corresponding to the discrete specter Coulomb functions. It means that in the Laguerre polynomial oscillations domain the aggregate contribution of the discrete specter accumulation point vicinity is defined only by main quantum number power decrease  and contains only oscillating (but not exponentially decreasing by sufficiently big arguments) functions. The corresponding contribution looks as
\be
\Psi_c^{acc}(\bz,\bq)\sim \mathop{\sum}\limits_{n=M}^\infty
\frac{(-1)^n}{n^4(B_2\ln n+B_1)}U(\by,\bq,-\hx)\times
\label{res-000}
\ee
$$
\times(8\pi n|\alpha_1|t_n)^{-1/2}
\left(\frac{|\alpha_1|t_n}{4}\right)^{-1/4}\left(1-\frac{|\alpha_1|t_n}{4}\right)^{-1/4}
\cos\left(2n\left[\arccos\sqrt{\frac{|\alpha_1|t_n}{4}}-\sqrt{\frac{|\alpha_1|t_n}{4}}
\sqrt{1-\frac{|\alpha_1|t_n}{4}}\right]\right)\equiv
$$
$$
\equiv \mathop{\sum}\limits_{n=M}^\infty \Theta(n),\ \ \ \ t_n=\frac{x}{n^2}.
$$

It explains the low convergence of many computational results connected with charged clusters scattering above the break-up threshold.

Calculate the asymptotics of the expression (\ref{res-000}) by $x\gg 1$ with the help of Poisson method. To do it, consider the generalized decomposition formula of the $\delta$-periodic function in the plane waves:
\be
\mathop{\sum}\limits_{n=-\infty}^\infty \delta(n-u)= \mathop{\sum}\limits_{l=-\infty}^\infty
e^{2\pi i lu}.
\label{puas-1}
\ee
Multiply the left and right-hand sides of the equation (\ref{puas-1})by the function $\Theta(u)$ and integrate over $du$ on the real axis. The function $\Theta$ was defined in the equation (\ref{res-000}) by integer arguments. In this case we consider $\Theta(u)$ as a continuous argument function defined by zero at $u<M-\varepsilon$,
$\ 0<\varepsilon<1$.

As an integration result we re-write the expression (\ref{res-000}) in the form of the integrals sum
$$
\Psi_c^{acc}(\bz,\bq)\sim \mathop{\sum}\limits_{l=-\infty}^\infty
\int_{M}^\infty e^{2\pi i l u}\Theta(u)du.
$$
Substituting the variable  $s=\frac14|\alpha_1|\frac{x}{u^2}$ and introducing the notations 
$R\equiv \sqrt{|\alpha_1|x}$, $\ d\equiv \frac{R^2}{4M^2}$, $\ 0<d<1$
we come to the following result
\be
\Psi_c^{acc}(\bz,\bq)\sim U(\by,\bq,-\hx)\frac{1}{2\sqrt{\pi}R^{3/2}}
\mathop{\sum}\limits_{l=-\infty}^\infty
\int_0^{d}ds \frac{ \chi_d^\delta(s) }{B_1+B_2\ln\left(\frac{R}{2\sqrt{s}}\right)}\times
\label{puas-2}
\ee
$$
\times
\frac{e^{i\pi\frac{R}{\sqrt{s}}(l+1/2)}}{(1-s)^{1/4}}
\cos\left(\frac{R}{\sqrt{s}}\left[\arccos(\sqrt{s})-\sqrt{s(1-s)}\right]\right)
$$
or, what is the same, 
\be
\Psi_c^{acc}(\bz,\bq)\sim U(\by,\bq,-\hx)\frac{1}{2\sqrt{\pi}R^{3/2}}
\mathop{\sum}\limits_{l=-\infty}^\infty
\frac12\int_0^{d} ds \frac{e^{iR\Phi_l^+(s)}+e^{iR\Phi_l^-(s)}}{B_1+B_2\ln\left(\frac{R}{2\sqrt{s}}\right)}
\frac{\chi_d^\delta(s)}{(1-s)^{1/4}}.
\label{puas-3}
\ee
We use here the notations
\be
\Phi_l^\pm(s)\equiv \frac{\pi}{\sqrt{s}}\left(l+\frac12\right)\pm\frac{1}{\sqrt{s}}
\left(\arccos(\sqrt{s})-\sqrt{s(1-s)}\right),
\label{phase-pm}
\ee
The function $\chi_d^\delta(s)$ is a patch function in the radius $\delta$ circle with the center in the point $s=d$ in the complex plane $s$. The exit from the real axis into the complex plane $s$ proves effective in further calculations. It is easy to show that the equation     
$$
{\Phi_l^\pm}'(s)=-\frac{1}{2s^{3/2}}\left\{
\frac{\pi}{2}+l\pi\pm \left(\arccos(\sqrt{s})+\sqrt{s(1-s)}\right)\right\}=0
$$
has no roots on the interval $0<s<d<1$ by no $l\in \bZ$. Thus the phase functions do not generate stationary points. 
On the other side integration by parts in the integrals contained in the expression (\ref{puas-3}) is not possible due to the presence of the logarithmic singularities of the denominator which after a differentiation become non-integrable.

Having used the identity
$$
-\frac{\pi}{2}+\arccos(\sqrt{s})-\sqrt{s(1-s)}=-\int_0^s\sqrt{\frac{1-\tau}{\tau}}d\tau,
$$
rewrite the expression  (\ref{phase-pm}) in the form of the integral presentations 
$$
\Phi_l^+=\frac{\pi(l+1)}{\sqrt{s}}-\frac{1}{\sqrt{s}}\int_0^s\sqrt{\frac{1-\tau}{\tau}}d\tau,
$$
$$
\Phi_l^-=\frac{\pi l}{\sqrt{s}}+\frac{1}{\sqrt{s}}\int_0^s\sqrt{\frac{1-\tau}{\tau}}d\tau.
$$
We are now going to evaluate the expression (\ref{puas-3}) asymptotically at large coordinates.
Rewrite the expression (\ref{puas-3}) in the following form 
\be
\Psi_c^{acc}(\bz,\bq)=\Psi_c^{I}(\bz,\bq)+\Psi_c^{II}(\bz,\bq)+\Psi_c^{III}(\bz,\bq)+\Psi_c^{IV}(\bz,\bq).
\label{puas-000}
\ee
We use here the notations 
\be
\Psi_c^{I}(\bz,\bq)= U(\by,\bq,-\hx)\frac{1}{2\sqrt{\pi}R^{3/2}}
\mathop{\sum}\limits_{l=0}^\infty
\frac12\int_0^{d} ds \frac{e^{iR\left[\frac{1}{\sqrt{s}}\int_0^s\sqrt{\frac{1-\tau}{\tau}}d\tau+\frac{\pi l}{\sqrt{s}}
\right]}}
{B_1+B_2\ln\left(\frac{R}{2\sqrt{s}}\right)}
\frac{\chi_d^\delta(s)}{(1-s)^{1/4}},
\label{puas-001}
\ee
\be
\Psi_c^{II}(\bz,\bq)= U(\by,\bq,-\hx)\frac{1}{2\sqrt{\pi}R^{3/2}}
\mathop{\sum}\limits_{l=-\infty}^{-1}
\frac12\int_0^{d} ds \frac{e^{-iR\left[\frac{1}{\sqrt{s}}\int_0^s\sqrt{\frac{1-\tau}{\tau}}d\tau-\frac{\pi(l+1)}{\sqrt{s}}
\right]}}
{B_1+B_2\ln\left(\frac{R}{2\sqrt{s}}\right)}
\frac{\chi_d^\delta(s)}{(1-s)^{1/4}},
\label{puas-002}
\ee
\be
\Psi_c^{III}(\bz,\bq)= U(\by,\bq,-\hx)\frac{1}{2\sqrt{\pi}R^{3/2}}
\mathop{\sum}\limits_{l=-\infty}^{-1}
\frac12\int_0^{d} ds \frac{e^{iR\left[\frac{1}{\sqrt{s}}\int_0^s\sqrt{\frac{1-\tau}{\tau}}d\tau+\frac{\pi l}{\sqrt{s}}
\right]}}
{B_1+B_2\ln\left(\frac{R}{2\sqrt{s}}\right)}
\frac{\chi_d^\delta(s)}{(1-s)^{1/4}},
\label{puas-003}
\ee
\be
\Psi_c^{IV}(\bz,\bq)= U(\by,\bq,-\hx)\frac{1}{2\sqrt{\pi}R^{3/2}}
\mathop{\sum}\limits_{l=0}^{\infty}
\frac12\int_0^{d} ds \frac{e^{-iR\left[\frac{1}{\sqrt{s}}\int_0^s\sqrt{\frac{1-\tau}{\tau}}d\tau-\frac{\pi(l+1)}{\sqrt{s}}
\right]}}
{B_1+B_2\ln\left(\frac{R}{2\sqrt{s}}\right)}
\frac{\chi_d^\delta(s)}{(1-s)^{1/4}}.
\label{puas-004}
\ee
Moving the integration contour to the lower semi-plane, we obtain that the main contribution into the expression $\Psi_c^{I}(\bz,\bq)$
is made by the term corresponding to $l=0$. This contribution is of the form
\be
\Psi_c^{I}(\bz,\bq)= U(\by,\bq,-\hx)\frac{1}{4\sqrt{\pi}R^{5/2}}\hat{\Upsilon}(R),
\label{final-001}
\ee
where где
\be
\hat{\Upsilon}(R)\equiv -ie^{2iR}\int_0^\infty\frac{dt e^{-t/3}}{C(R)-D\ln t}
\left(1+O\left(\frac{1}{R}\right)\right).
\label{final-001-1}
\ee
Here the notations are used 
\be
C(R)=B_1+\frac32 B_2\ln R-B_2\ln 2+iB_2\frac{\pi}{4},\ \ \ \ \ \ \ D=\frac12 B_2.
\label{crc-00}
\ee
The main order of the expression  (\ref{final-001}) can be written in the form 
$$
\Psi_c^{I}(\bz,\bq)= U(\by,\bq,-\hx)\frac{-3i}{4\sqrt{\pi}R^{5/2}}
\frac{e^{2iR}}{C(R)}\left(1+O\left(\frac{1}{\ln R}\right)\right).
$$

Moving the integration contour to the upper semi-plane we obtain that the main contribution into the expression 
 $\Psi_c^{II}(\bz,\bq)$
is made by the term corresponding to $l=-1$. The contribution is of the form 
\be
\Psi_c^{II}(\bz,\bq)= U(\by,\bq,-\hx)\frac{1}{4\sqrt{\pi}R^{5/2}}\hat{\Upsilon}^*(R).
\label{final-002}
\ee
The main order of the expression (\ref{final-002}) can be written in the form 
$$
\Psi_c^{II}(\bz,\bq)= U(\by,\bq,-\hx)\frac{3i}{4\sqrt{\pi}R^{5/2}}
\frac{e^{-2iR}}{C^*(R)}\left(1+O\left(\frac{1}{\ln R}\right)\right).
$$
Analogously we can show that the contributions into the asymptotics of the expressions 
 $\Psi_c^{III}(\bz,\bq)$ and $\Psi_c^{IV}(\bz,\bq)$
are exponentially small. 

It allows us to state that the total contribution into the continuous spectrum three-body eigenfunction asymptotics of the pairwise Coulomb excitations is of the form
\be
\Psi_c^{acc}(\bz,\bq)=\frac{3}{2\sqrt{\pi}}U(\by,\bq,-\hx)
\frac{\sin(2R)}{C(R)R^{5/2}}\left(1+O\left(\frac{1}{\ln R}\right)\right),\ \ \ \ R=\sqrt{|\alpha_1|x}
\label{result-00}
\ee
in the domain bounded by the relation  $1\ll x\le \frac{4M^2}{|\alpha_1|},\ \ M\gg 1$.

It is this result that is central in this work. Formulate the obtained result in the following way.

\vskip0.5cm
{\bf Statement 1:}

In the domains of the type (\ref{sep-j}) of configuration space where true is "an almost separation of variables" in the vicinities of the screens 
 $\sigma$ by  $y >> x$ (where $\by$ and  $\bx$ are Jacobi variables corresponding to the pairwise subsystem with a Coulomb attraction potential) the total contribution of the discrete spectrum pairwise states into the continuous spectrum three-body eigenfunction is of the form:
\be
\Psi^{D}(\bz,\bq)=
\mathop{\sum}\limits_{n=1}^M\mathop{\sum}\limits_{l=0}^{n-1}\mathop{\sum}\limits_{m=-l}^l
\int_{\bR^3}d\bp'\psi_{nlm}(x)Y_l^m(\hx)\psi_c^{eff}(\by,\bp')\delta({p'}^2-\frac{|\alpha_1|^2}{4n^2}-E)
R_{nlm}(\bq,\bp')\ +\  \Xi^{total}(\bz,\bq),
\label{sep-chi-000}
\ee
where $M\gg 1$  is a main quantum number in the pairwise Coulomb subsystem, while 
 $\Xi^{total}(\bz,\bq)$ is a total contribution of all discrete spectrum pairwise states with a main quantum number larger than 
 $M$.

Then the expression for  $\Xi^{total}(\bz,\bq)$ is of the form 
\be
\Xi^{total}(\bz,\bq)
=\frac{3}{2\sqrt{\pi}}U(\by,\bq,-\hx)
\frac{\sin(2\sqrt{|\alpha_1|x})}{C(\sqrt{|\alpha_1|x})(|\alpha_1|x)^{5/4}}\left(1+O\left(\frac{1}{\ln(|\alpha_1|x)}\right)\right),
\label{result-00-0}
\ee
$$
C(R)=B_1+\frac32 B_2\ln R-B_2\ln 2+iB_2\frac{\pi}{4},\ \ \ \ \ \ \ D=\frac12 B_2
$$
in the domain bounded by the relation  $1\ll x\le \frac{4M^2}{|\alpha_1|},\ \ M\gg 1$.
The function  $U(\by,\bq,-\hx)$ is a smooth function of its arguments, the real constants 
 $B_1,\ B_2$ were defined above. The numerical calculations gives the result: 
$$
B1=-0.310±0.001,\ \ \ B2=-0.63±0.001.
$$ 

Note that the more pairwise states will be counted explicitly in the total (\ref{sep-chi-000}), i.e. with a rise of parameter $M$, the wider will be the application domain of the expression (\ref{result-00-0}). At that the expression (\ref{result-00-0}) itself proves to be the main order of asymptotics at big values of coordinate $x$. Thus the asymptotics application domain with a rise of $M$ will be moving to the domain of big values of $x$. On the other side the asymptotics structure of (\ref{result-00-0}) does not explicitly depend on the row $M$ truncation  parameter and is, therefore, a full coordinate asymptotics with respect to variable $x$ of the wave function at infinity. 

Note as well the oscillatory character and slow power decrease at infinity of the obtained expression which makes it necessary to count the given asymptotic contribution against the background of the exponentially decreasing explicitly counted terms of the sum (\ref{sep-chi-000}).

Thus true is 

{\bf Statement 2:}

The total contribution of pairwise Coulomb highly excited states into the scattering state asymptotics in the problem of three charged quantum particles is of the following structure
$$
\Psi_c(\bz,\bq)= U(\by,\bq,-\hx)\frac{1}{4\sqrt{\pi}R^{5/2}}
\left(\hat{\Upsilon}(R)+{\hat{\Upsilon}}^*(R)\right),\ \ \ \ \ \ \ \ R=\sqrt{|\alpha_1|x},
$$
where the function $\hat{\Upsilon}(R)$ is described in the equation  (\ref{final-001-1}).
This statement is the central one in the work.

\newpage

\section{Appendix A. Definition of the Kernel $g_n(\bx,\homega)$ of the Generating Integral.}

Consider continuous specter eigenfunction $\tilde{\psi}_c(\bx,\bk)$ defined in the standard way with an accuracy down to the normalization (the normalizing coefficient is set equal to a unit):
\be
\tilde{\psi}_c(\bx,\bk)=e^{i\langle\bk,\bx\rangle}\Phi(-i\gamma,1,ikx(1-\langle\hk,\hx\rangle))
\label{a1}
\ee
of Coulomb two-body Schrodinger operator (here $N_0(k)$ is a normalizing constant) and make its partial analysis. 
Use the Kummer (9.212.1) \cite{Grad} transformation
$$
\Phi(a,c,z)=e^z\Phi(c-a,c,-z),
$$
which makes it possible to rewrite (\ref{a1}) in the form of 
\be
\tilde{\psi}_c(\bx,\bk)=e^{ikx}\Phi(1+i\gamma,1,-ikx(1-\langle\hk,\hx\rangle)).
\label{a2}
\ee
Now the angular dependence is concentrated only in the confluent hypergeometric function.
Expand it in an orthogonal Legendre polynomials $P_l$ series. According to, for example,
(8.904) \cite{Grad} the decomposition takes form
\be
\Phi(1+i\gamma,1,-ikx(1-t))=\mathop{\sum}\limits_{l=0}^\infty(2l+1)\Phi_l(k,x)P_l(t).
\label{a3}
\ee
Here the notations are introduced:  $t=\langle\hk,\hx\rangle$,
$\Phi_l(k,x)$ are partial components of the function  $\Phi(1+i\gamma,1,-ikx(1-t))$,
which according to the Legendre polynomials orthogonality conditions are calculated in the following way
\be
\Phi_l(k,x)=\frac12\int_{-1}^1 dt\Phi(1+i\gamma,1,-ikx(1-t))P_l(t).
\label{a4}
\ee
To calculate the integral in (\ref{a4}) we will use a hypergeometric decomposition
\be
\Phi(a,1,b(1-t))=\mathop{\sum}\limits_{j=0}^{\infty} \frac{(a)_j}{(j!)^2}b^j(1-t)^j,\ \ \ \
a=1+i\gamma,\ \ \ b=-ikx,\ \ \ (a)_j=\frac{\Gamma(a+j)}{\Gamma(a)},
\label{hyper-1}
\ee
in the terms of which
\be
\Phi_l(k,x)=\frac12\mathop{\sum}\limits_{j=0}^{\infty} \frac{(a)_j}{(j!)^2}b^j
\int_{-1}^1 dt(1-t)^jP_l(t).
\label{a5}
\ee
Making a change of the variable $s^2=\frac{1-t}{2}$ in the integral, calculate it explicitly and obtain a new hypergeometric decomposition leading to the following finite relation for a partial component

\be
\Phi_l(k,x)=\frac{(-1)^l}{\Gamma(l+2)}\mathop{\lim}\limits_{u\rightarrow -m}
\frac{1}{\Gamma(u)}{_2F_2}(a,1;u,l+2;2b),\ \ \ m=l-1.
\label{f22}
\ee
Now use the equation
\be
\mathop{\lim}\limits_{u\rightarrow -m}\frac{1}{\Gamma(u)}{_2F_2}(A,B;u,D;z)=
\frac{(A)_{m+1}(B)_{m+1}}{(D)_{m+1}}\frac{z^{m+1}}{(m+1)!}
{_2F_2}(A+m+1,B+m+1;m+2,D+m+1;z),
\label{f-fin}
\ee
where 
$$
A=a,\ \ B=1,\ \ m=l-1,\ \ D=l+2,\ \ z=2b.
$$
Substituting the equation (\ref{f-fin}) into the в (\ref{f22}), obtain the final expression for the partial component
\be
\Phi_l(k,x)=\frac{\Gamma(i\gamma+l+1)}{\Gamma{i\gamma+1}\Gamma(2l+2)}(2ikx)^l
\Phi(i\gamma+l+1,2l+2,-2ikx).
\label{f-fin1}
\ee
Substituting the obtained relation into the decomposition (\ref{a2})-(\ref{a3}) for a Schrodinger operator two-body Coulomb continuous spectrum eigenfunction we obtain
\be
\tilde{\psi}_c(\bx,\bk)=e^{ikx}\mathop{\sum}\limits_{l=0}^{\infty}(2l+1)
\frac{\Gamma(i\gamma+l+1)}{\Gamma(i\gamma+1)\Gamma(2l+2)}
(2ikx)^l\Phi(i\gamma+l+1,2l+2,-2ikx) P_l(t),\ \ \ \ t=\langle\hk,\hx\rangle.
\label{f-0}
\ee
Consider now the analytical extension of the function constructed in the upper half/semi- plane of the complex plane $k$ by
$$
k=k_n= i\frac{|\alpha|}{2n},\ \ \ \ \alpha<0,\ \ \
i\gamma=i\frac{\alpha}{2k}|_{k=k_n}=i\frac{\alpha}{2k_n}=-n,\ \ \  n=1,2,3,\dots.
$$
In this case the equation (\ref{f-0}) takes form 
\be
\tilde{\psi}_c(\bx,\bk_n)=e^{-\frac{|\alpha|}{2n}x}\mathop{\sum}\limits_{l=0}^{\infty}\beta_{nl}
\frac{2l+1}{\Gamma(2l+1)}
(-\frac{|\alpha|}{n}x)^l\Phi\left(-n+l+1,2l+2,\frac{|\alpha|}{n}x\right) P_l(t),\ \ \ \ t=\langle\hk,\hx\rangle.
\label{f-1}
\ee
We consider here the vector $\bk_n$ as a vector with a direction coinciding with the direction of the initial vector
$\bk\ $ ($\hk_n=\hk$) and the length taking purely imaginary value in the upper half-plane of the complex plane  $k\ $ ($k_n=i\frac{|\alpha|}{2n}$).
Here the coefficient  $\beta_{nl}$ is defined as follows 
\be
\beta_{nl}=(1-n)(2-n)\dots(l-n).
\label{beta-0}
\ee
Note that according to the equation (\ref{beta-0}),the coefficient $\beta_{nl}$ becomes zero by $l\ge n$. Thus the expression (\ref{f-1}) proves to be the finite sum of terms and can be written in the form of
\be
\tilde{\psi}_c(\bx,\bk_n)=e^{-\frac{|\alpha|}{2n}x}
\Phi\left(1-n,1,\frac{|\alpha|}{2n}x(1-\langle\hk,\hx\rangle)\right)=
\label{f-2}
\ee
$$
=4\pi e^{-\frac{|\alpha|}{2n}x}\mathop{\sum}\limits_{l=0}^{n-1}\mathop{\sum}\limits_{m=-l}^{l}
\beta_{nl}
\frac{1}{(2l+1)!}
(-\frac{|\alpha|}{n})^l x^l\Phi\left(-n+l+1,2l+2,\frac{|\alpha|}{n}x\right) Y_l^m(\hx)
Y_l^m(\hk),
$$
where $Y_l^m$ are corresponding spherical functions.

Finally introduce the function 
\be
g_n(\bx,\hk)\equiv \tilde{\psi}_c(\bx,\bk_n)
\label{gn-def}
\ee
according to the equation (\ref{f-2}). We presented a function, which partial components with an accuracy down to a normalization coincide with discrete spectrum Coulomb eigenfunctions by the fixed main quantum number $n$.

\newpage

\section{Appendix B. Calculation of the Angular Part of the Normalization Integral.}

In the section we calculate an angular integral in the normalization expression (\ref{normi-1})
 \be
\Sigma_n(\hk',\hk'')\sim
\frac{2^6 n^3}{\pi\sqrt{\upsilon}|\alpha_1|^3}\int_{\Delta}^{\upsilon-\Delta} dt t^{3/2}
\int_{\bS^2}d\hx
e^{-t\left(1-\frac12(\sin^2\frac{\theta}{2}+ \sin^2\frac{\tilde{\theta}}{2} )\right)}\times
\label{int-b10}
\ee
$$
\times\frac{1}{\sqrt{\sin\frac{\theta}{2} \sin\frac{\tilde{\theta}}{2}}}
\left(1-\frac{t}{\upsilon}\sin^2\frac{\theta}{2}\right)^{-1/4}
\left(1-\frac{t}{\upsilon}\sin^2\frac{\tilde{\theta}}{2}\right)^{-1/4}\times
$$
$$
\times
\left\{\cos\left(\frac{\upsilon}{4}\left(2(\theta_*-\tilde{\theta}_*)-
(\sin 2\theta_*-\sin 2\tilde{\theta}_*)\right)\right)+
\sin\left(\frac{\upsilon}{4}\left(2(\theta_*+\tilde{\theta}_*)-
(\sin 2\theta_*+\sin 2\tilde{\theta}_*)\right)\right)\right\}.
$$
Here the parameter $\Delta$ in the limits of integration is connected with defining the limits of the oscillations domain:
$\Delta=O(\upsilon^\varrho),\ \ 0<\varrho<\frac13$.
In the Cartesian coordinate system $(x,y,z)$ the unit vector $\hk''$ is directed along the axis $z$, the unit vector $\hx$  is characterized by the pair of angles $(\theta_{k'},\ \varphi_{k'})$.
At that 
\be
\cos\tilde{\theta}=\sin\theta\sin\theta_{k'}\cos(\varphi-\varphi_{k'})
+\cos\theta\cos\theta_{k'},
\label{connect-b}
\ee
\be
0\le\theta\le\pi,\ \ \ 0\le\theta_{k'}\le\pi,\ \ \ 0\le\varphi\le 2\pi,\ \ \ 0\le\varphi_{k'}\le 2\pi,
\label{domain}
\ee

After a change of the variable $s=\frac{t}{\upsilon}$ the expression (\ref{int-10}) takes form
 \be
\Sigma_n(\hk',\hk'')\sim
\frac{2^4 n^5}{\pi|\alpha_1|^3}
\int_{\bS^2}d\hx\int_{\delta}^{1-\delta} ds s^{3/2}
e^{-4ns\left(1-\frac12(\sin^2\frac{\theta}{2}+ \sin^2\frac{\tilde{\theta}}{2} )\right)}\times
\label{int-11}
\ee
$$
\times\frac{1}{\sqrt{\sin\frac{\theta}{2} \sin\frac{\tilde{\theta}}{2}}}
\left(1-s\sin^2\frac{\theta}{2}\right)^{-1/4}
\left(1-s\sin^2\frac{\tilde{\theta}}{2}\right)^{-1/4}\times
$$
$$
\times
\left\{\cos\left(n(2(\theta_*-\tilde{\theta}_*)-
(\sin 2\theta_*-\sin 2\tilde{\theta}_*))\right)+
\sin\left(n(2(\theta_*+\tilde{\theta}_*)-
(\sin 2\theta_*+\sin 2\tilde{\theta}_*))\right)\right\}.
$$
Here 
$$
\theta_*=\arccos\left(\sqrt{s}\sin\frac{\theta}{2}\right),\ \ \ \
\tilde{\theta}_*=\arccos\left(\sqrt{s}\sin\frac{\tilde{\theta}}{2}\right).
$$
Hereafter we will neglect the small parameter $\delta$ in the limits of integration on the variable
$s$,
$$
\delta=O\left(\frac{1}{n^{1-\varrho}}\right)\mathop{\rightarrow}\limits_{n\rightarrow \infty}0,\ \ \ 0<\varrho<\frac13,
$$
being interested in the normalizing integral behaviour only in the main order on $n$.

We are interested in the behaviour of the integral of the form
\be
F(\lambda)=\int_{\bS^2}d\hx f(\hx)e^{i\lambda S(\hx)},\ \ \ \ \lambda\gg 1.
\label{fed-1}
\ee
With the two-dimensional stationary point $\hx^0=(\theta_0,\varphi_0)$ the main contribution into the integral of the form (\ref{fed-1})
according to \cite{fed-1} is given by the expression
\be
F(\lambda)=\frac{2\pi}{\lambda}\exp\left\{i\lambda S(\hx^0)+\frac{i\pi}{4} sgn S''_{xx}(\hx^0)\right\}
\left|\det S''_{xx}(\hx^0)\right|^{-\frac12}\left(f(\hx^0)+O(\lambda^{-1})\right),
\label{fedor-5}
\ee
where $S''_{xx}(\hx^0)$ is a second derivatives matrix in the stationary phase point, while $sgn S''_{xx}(\hx^0)$ is a difference
of the number of positive and negative eigenvalues of the matrix in the same point.

As was stated above, the big parameter defining the angular variable integral behavior in the expression (\ref{int-11})
is effectively contained only in oscillating functions. We will make total calculations only for a summand in the subintegral
expression in (\ref{int-11}) containing a sinus. Looking ahead, say that making similar calculations for a summand containing a cosine,
it is easy to see that this summand either does not generate a stationary phase point, or generates an a priori smaller contribution into the expression (\ref{int-11}).

Thus, consider the sinus argument in the expression (\ref{int-11}) being a big parameter multiplier or, in other words, a phase function
\be
S(\theta,\varphi)=2\arccos\left(\sqrt{s}\sin\frac{\theta}{2}\right)+
2\arccos\left(\sqrt{s}\sin\frac{\tilde{\theta}}{2}\right)-
\label{phase}
\ee
$$
-2\sqrt{s}\sqrt{1-s\sin^2\frac{\theta}{2}}\sin\frac{\theta}{2}-
2\sqrt{s}\sqrt{1-s\sin^2\frac{\tilde{\theta}}{2}}\sin\frac{\tilde{\theta}}{2}
$$
upon the condition of constraint (\ref{connect-b}).

The existence of the stationary phase point $(\theta_0,\varphi_0)$ is connected with a solution of the equations system
\be
\left\{
\begin{tabular}{l}
$\frac{\partial S(\theta,\varphi)}{\partial\varphi}|_{(\theta_0,\varphi_0)} =0,$ \\
$\frac{\partial S(\theta,\varphi)}{\partial\theta}|_{(\theta_0,\varphi_0)} =0.$ \\
\end{tabular}
\right.
\label{eq-1}
\ee
The first of the equations (\ref{eq-1}) takes form 
\be
2\sqrt{s}\sqrt{1-s\sin^2\frac{\tilde{\theta}}{2}}\
\frac{\partial\sin\frac{\tilde{\theta}}{2}}{\partial\varphi}=0,
\ee
where $s\in[0,1]$ is an external parameter. The solution of the equation  
$\frac{\partial\sin\frac{\tilde{\theta}}{2}}{\partial\varphi}=0$ in accordance with the condition of constraint
(\ref{connect-b}) gives 
$$
\sin(\varphi-\varphi_{k'})=0
$$
or finally two different solutions
\be
 \varphi=\varphi_{k'},\ \mbox{или} \ \varphi=\varphi_{k'}+\pi.
 \label{phi-1-2}
 \ee
Thus according to (\ref{connect-b}) obtain 
\be
\cos\tilde{\theta}=\cos(\theta-\theta_{k'})\ \ \mbox{or} \ \
\cos\tilde{\theta}=\cos(\theta+\theta_{k'})
\label{stat-1}
 \ee
 At that the minus sign corresponds to the first of the conditions (\ref{phi-1-2}), 
 while the plus sign corresponds to the second condition.

The second of the conditions (\ref{eq-1})can be reduced to the form
\be
\cos\frac{\theta}{2}
\sqrt{1-s\sin^2\frac{\theta}{2}} +
\cos\frac{\tilde{\theta}}{2}
\sqrt{1-s\sin^2\frac{\tilde{\theta}}{2}} = 0.
\ee
The solution of the equation is structured as follows
\be
\frac{\tilde{\theta}}{2}=\pi-\frac{\theta}{2}.
\label{eq-2}
\ee
The solution together with the first of the conditions (\ref{stat-1})
is reduced to the form
\be
\theta=\pi+\frac{\theta_{k'}}{2}.
\label{df-1}
\ee
The solution contradicts to the conditions (\ref{domain}) on the angular variables definition domain.
The term (\ref{eq-2}) together with the second of the conditions  (\ref{stat-1})
is reduced to the form 
\be
\theta=\pi-\frac{\theta_{k'}}{2}.
\label{df-2}
\ee
The term satisfies the conditions (\ref{domain}) and defines the two-dimensional stationary point
$(\theta_0,\varphi_0)$:
\be
\theta_0=\pi-\frac{\theta_{k'}}{2},\ \ \ \varphi_0=\varphi_{k'}+\pi.
\label{stat-00}
\ee
According to, for example, \cite{fed-1}, define the second derivative matrix of the phase function. The matrix elements are of the form
$$
\frac{\partial^2S}{\partial\varphi^2}|_{(\theta_0,\varphi_0)}=\sqrt{s}\
 \sin\theta_{k'}\sin\frac{\theta_{k'}}{2}
 \frac{\sqrt{1-s\cos^2\frac{\theta_{k'}}{4}}}{\cos\frac{\theta_{k'}}{4}},
 $$
$$
\frac{\partial^2S}{\partial\varphi\partial\theta}|_{(\theta_0,\varphi_0)}=0,
$$
$$
\frac{\partial^2S}{\partial\theta^2}|_{(\theta_0,\varphi_0)}=
\frac{2\sqrt{s}\cos\frac{\theta_{k'}}{4}}
{\sqrt{1-s\cos^2\frac{\theta_{k'}}{4}}}
\left(1-s\cos\frac{\theta_{k'}}{2}\right).
$$
The second derivative matrix determinant is of the form
$$
\det (S''_{xx})|_{(\theta_0,\varphi_0)}=2s\sin\theta_{k'}\sin\frac{\theta_{k'}}{2}
\left(1-s\cos\frac{\theta_{k'}}{2}\right).
$$
Finally the difference of the positive and negative eigenvalues of the second derivatives matrix in 
the stationary phase point is
$$
sgn S''_{xx}(\theta_0,\varphi_0)=\nu_+-\nu_-=2.
$$
 In accordance with the expression (\ref{fedor-5}) the equation (\ref{int-11}) takes form 
\be
\Sigma_n(\hk',\hk'')\sim
\Im\left\{\frac{2^4 n^4}{|\alpha_1|^3}
\int_{0}^{1} ds s
e^{-4ns\sin^2\frac{\theta_{k'}}{4}}
\frac{e^{inS(\theta_0,\varphi_0)+i\frac{\pi}{2}}}
{\cos\frac{\theta_{k'}}{4}\sqrt{\cos\frac{\theta_{k'}}{2}}
\sqrt{1-s\cos^2\frac{\theta_{k'}}{4}}\sqrt{1-s\cos\frac{\theta_{k'}}{2}}}
\right\}.
\label{int-12}
\ee

\newpage

\section{Appendix C. The Dependence of the Normalizing Integral on the Big Parameter $n$.}

In the section we will consider the behaviour of the dual integral of the form
\be
\Omega_n=\int_0^1 du \int_0^1 dv e^{-4n\beta uv f(u,v)}e^{4inu^{3/2}g(u,v)}F(u,v).
\label{uv-11}
\ee
as a large parameter $n\gg 1$ function. Here
$$
u=\rho^2,\ \ \ \ \ v=\sin^2\omega,\ \ \ \ \beta=\frac{1}{16}.
$$
We also used the notations 
\be
f(u,v)=(1+\phi(u,v))(1-\beta u(1-v))\ge 0,\ \ \ \phi(u,v)\mathop{=}\limits_{u,v\rightarrow 0}O(uv),
\label{def-f0}
\ee
\be
g(u,v)=\frac{13}{6}+v(1-v)+ \sigma(u,v),\ \ \ \sigma(u,v)\mathop{=}\limits_{u\rightarrow 0}O(u^{1/2}).
\label{def-g0}
\ee
The functions $F(u,v)$,  $\phi(u,v)$ and $\sigma(u,v)$ are bounded, twice continuously differentiable functions of their arguments.

As the first step, split the integral (\ref{uv-11} into two parts:
\be
\Omega_n=\Omega_n^I+\Omega_n^{II},
\label{dif-1}
\ee
where
\be
\Omega_n^I=\int_0^1 du \int_0^1 dv e^{-4n\beta uv f(u,v)}e^{4inu^{3/2}g(u,v)}(F(u,v)-F(0,v)),
\label{dif-11}
\ee
\be
\Omega_n^{II}=\int_0^1 du \int_0^1 dv e^{-4n\beta uv f(u,v)}e^{4inu^{3/2}g(u,v)}F(0,v).
\label{dif-12}
\ee
The expression under the integral sign in (\ref{dif-11}) becomes zero in the point $u=0$. We will integrate the expression by parts on variable $v$.
\be
\Omega_n^I=\frac{1}{4n}\int_0^1 du \int_0^1 dv \left(e^{-4n\beta uv f(u,v)+4inu^{3/2}g(u,v)}\right)'_v
\frac{F(u,v)-F(0,v)}{-\beta u(vf)'_v+iu^{3/2}g'_v}=
\label{dif-13}
\ee
$$
=\frac{1}{4n}\int_0^1 du e^{-4n\beta u f(u,1)+4inu^{3/2}g(u,1)}
\frac{F(u,1)-F(0,1)}{[-\beta u(vf)'_v+iu^{3/2}g'_v]_{v=1}}-
$$
$$
-\frac{1}{4n}\int_0^1 du e^{4inu^{3/2}g(u,0)}
\frac{F(u,0)-F(0,0)}{[-\beta u(vf)'_v+iu^{3/2}g'_v]_{v=0}}+
$$
$$
+\frac{1}{4n}\int_0^1 du \int_0^1 dv e^{-4n\beta uv f(u,v)+4inu^{3/2}g(u,v)}
\frac{\partial}{\partial v}\left[\frac{F(u,v)-F(0,v)}{\beta u(vf)'_v-iu^{3/2}g'_v}\right].
$$
Note that the expression in the fourth line of the equation (\ref{dif-13}) is of the order $O(n^{-2})$,
as it allows a repeated procedure of integrating by parts. The integrals in the second and third lines are a priori of the order $o(n^{-1})$, which is defined by the saddle-point method or the stationary phase method.

\subsection{The Description of the Behavior of the Contribution $\Omega_n^{II}$}

We will study the behavior of the integral $\Omega_n^{II}$ (\ref{dif-12}).
Introduce the notation 
\be
G(u,v)\equiv \beta v f(u,v)-i\sqrt{u}g(u,v).
\label{def-g}
\ee
and separate a small vicinity of zero on variable $u$, having substituted the variable $t=nu$:
$$
\Omega_n^{II}=\frac{1}{n}\int_0^1 dt\int_0^1dv e^{-4tG(0,v)}F(0,v)+
\int_{\frac{1}{n}}^1du\int_0^1 dv e^{-4nuG(u,v)}F(0,v).
$$
The integral in the first term does not depend on the large parameter. The integral in the second term will be integrated by parts on variable $v$. Thus,
$$
\Omega_n^{II}=\frac{1}{n}\varpi_1+\int_{\frac{1}{n}}^1du\int_0^1 dv \left(e^{-4nuG(u,v)}\right)'_v
\frac{F(0,v)}{-4nuG'_v(u,v)}.
$$
Here 
\be
\varpi_1\equiv \int_0^1 dt\int_0^1dv e^{-4tG(0,v)}F(0,v).
\label{alpha-1}
\ee
Integration by parts leads to the following result:
\be
\Omega_n^{II}=\frac{1}{n}\alpha_1-\frac{1}{4n}F(0,1)\int_{\frac1n}^1\frac{du}{u}
\frac{e^{-4nuG(u,1)}}{G'_v(u,1)}+
\frac{1}{4n}F(0,0)\int_{\frac1n}^1\frac{du}{u}
\frac{e^{-4nuG(u,0)}}{G'_v(u,0)}+
\label{dif-31}
\ee
$$
+\frac{1}{4n}\int_{\frac1n}^1\frac{du}{u}\int_0^1dve^{-4nuG(u,v)}\frac{\partial}{\partial v}
\left(\frac{F(0,v)}{G'_v(u,v)}\right).
$$
For each of the integrals obtained in the equation (\ref{dif-31}) we will introduce its own notation:
\be
\Omega_n^{II}=\frac{1}{n}\varpi_1+\Omega_n^{III}+\Omega_n^{IV}+\Omega_n^{V},
\label{dif-32}
\ee
where the constant  $\varpi_1$ is described in the expression  (\ref{alpha-1}),
\be
\Omega_n^{III}=-\frac{1}{4n}F(0,1)\int_{\frac1n}^1\frac{du}{u}
\frac{e^{-4nuG(u,1)}}{G'_v(u,1)},
\label{III-0}
\ee
\be
\Omega_n^{IV}=\frac{1}{4n}F(0,0)\int_{\frac1n}^1\frac{du}{u}
\frac{e^{-4nuG(u,0)}}{G'_v(u,0)},
\label{IV-0}
\ee
\be
\Omega_n^{V}=\frac{1}{4n}\int_{\frac1n}^1\frac{du}{u}\int_0^1dve^{-4nuG(u,v)}\frac{\partial}{\partial v}
\left(\frac{F(0,v)}{G'_v(u,v)}\right).
\label{V-0}
\ee

Start our consideration with the expression $\Omega_n^{III}$.

\subsection{The Description of the Behaviour of the Contribution $\Omega_n^{III}$}
Consider the expression
\be
\Omega_n^{III}=-\frac{1}{4n}F(0,1)\int_{\frac1n}^1\frac{du}{u}
\frac{e^{-4nuG(u,1)}}{G'_v(u,1)}.
\label{dif-300}
\ee
We will integrate by parts on variable $u$, eliminating the singularity in zero:
$$
\Omega_n^{III}=-\frac{1}{4n}F(0,1)\int_{\frac1n}^1 du\ln'u
\frac{e^{-4nuG(u,1)}}{G'_v(u,1)}=
$$
$$
=-\frac{1}{4n}F(0,1)\left[\ln n\frac{e^{-4G(0,1)}}{G'_v(0,1)}-
\int_{\frac1n}^1 du\ln u\left(\frac{e^{-4nuG(u,1)}}{G'_v(u,1)}\right)'_u\right]=
$$
$$
=-\frac{\ln n}{4n}F(0,1)\frac{e^{-4G(0,1)}}{G'_v(0,1)}-
F(0,1)\int_{\frac1n}^1 du\ln u\frac{G(u,1)+uG'_u(u,1)}{G'_v(u,1)}e^{-4nuG(u,1)}+
o\left(\frac{1}{n}\right).
$$
In the last integral we substitute the variable $t=un$. It leads to the following result
$$
\Omega_n^{III}=-\frac{\ln n}{4n}F(0,1)\frac{e^{-4G(0,1)}}{G'_v(0,1)}-
\frac{1}{n}F(0,1)\frac{G(0,1)}{G'_v(0,1)}\int_1^\infty dt(\ln t-\ln n)e^{-4tG(0,1)}
+o\left(\frac{1}{n}\right).
$$
We have used here the positivity $G(0,1)>0$ (\ref{def-g}), (\ref{def-f0})-(\ref{def-g0}) and therefore the exponential decrease of the expression under the integral sign. In its turn it made it possible to substitute the upper limit for infinity. The second term in the expression under the integral sign is computed explicitly and reduces the term of higher exponent of the order
$O\left(\frac{\ln n}{n}\right)$:
\be
\Omega_n^{III}=-\frac{1}{n}F(0,1)\frac{G(0,1)}{G'_v(0,1)}\int_1^\infty dt\ln t e^{-4tG(0,1)}
+o\left(\frac{1}{n}\right).
\label{dif-301}
\ee
Finally 
\be
\Omega_n^{III}=-\frac{1}{n}\varpi_2+o\left(\frac{1}{n}\right),
\label{dif-302}
\ee
where 
\be
\varpi_2\equiv F(0,1)\frac{G(0,1)}{G'_v(0,1)}\int_1^\infty dt\ln t e^{-4tG(0,1)}.
\label{alpha-2}
\ee

\subsection{The Description of the Behaviour of the Contribution  $\Omega_n^{IV}$.}

We will consider the expression $\Omega_n^{IV}$, defined in the equations (\ref{dif-31})-(\ref{dif-32}).
\be
\Omega_n^{IV}=\frac{1}{4n}F(0,0)\int_{\frac1n}^1\frac{du}{u}
\frac{e^{-4nuG(u,0)}}{G'_v(u,0)}
\label{def-iv}
\ee
Note that the expression $G(u,0)$,according to (\ref{def-g}), (\ref{def-f0})-(\ref{def-g0}),
is purely imaginary. Therefore the expression under the integral sign in $\Omega_n^{IV}$, contrary to the integral in
$\Omega_n^{III}$, contains a non-decreasing but just oscillating exponent.
At that the  main order of the behaviour of $G$ as a function of variable $u$ also changes, which, as we will 
demonstrate, proves to be essential for a logarithmic dependence on the large parameter $n$ to appear.

Thus, according to (\ref{def-g}), (\ref{def-g0}),
$$
G(u,0)=-i\sqrt{u}g(u,0),\ \ \ \ g(u,0)=\frac{13}{6}+O(u^{1/2}).
$$
Introduce the variable $t=u^{3/2}$. In its terms 
\be
\Omega_n^{IV}=\frac{1}{6n}F(0,0)\int_{\frac{1}{n^{3/2}}}^1\frac{dt}{t}
\frac{e^{4int\tilde{g}(t,0)}}{\tilde{G}'_v(t,0)}.
\label{def-iv-1}
\ee
The change of variables  $\rho=nt$ leads to the expression 
$$
\Omega_n^{IV}=\frac{1}{6n}\frac{F(0,0)}{\tilde{G}'_v(0,0)} \int_{\frac{1}{n^{1/2}}}^\infty\frac{d\rho}{\rho}e^{4i\rho\tilde{g}(0,0)}+o\left(\frac{1}{n}\right).
$$
We will split the obtained integral into two terms
\be
\frac{1}{6n}\frac{F(0,0)}{\tilde{G}'_v(0,0)}\int_{\frac{1}{n^{1/2}}}^\infty\frac{d\rho}{\rho}
e^{4i\rho\tilde{g}(0,0)}=
\frac{1}{n}\varpi_3+\frac{1}{6n}\frac{F(0,0)}{\tilde{G}'_v(0,0)}
\int_{\frac{1}{n^{1/2}}}^1\frac{d\rho}{\rho}
e^{4i\rho\tilde{g}(0,0)}.
\label{def-iv-2}
\ee
Here the notation is used 
\be
\varpi_3\equiv\frac{1}{6}\frac{F(0,0)}{\tilde{G}'_v(0,0)}\int_1^\infty\frac{d\rho}{\rho}
e^{4i\rho\tilde{g}(0,0)}.
\label{def-alpha3}
\ee
Once again integrating by parts the expression in the right-hand side of the equation (\ref{def-iv-2}),
we obtain
\be
\frac{1}{6n}\frac{F(0,0)}{\tilde{G}'_v(0,0)}
\int_{\frac{1}{n^{1/2}}}^1\frac{d\rho}{\rho}
e^{4i\rho\tilde{g}(0,0)}=
\frac{\ln n}{12n}\frac{F(0,0)}{\tilde{G}'_v(0,0)}
-i\frac{2}{3n}\frac{F(0,0)}{\tilde{G}'_v(0,0)}
\int_0^1 d\rho\ln\rho
e^{4i\rho\tilde{g}(0,0)}+o\left(\frac1n\right)=
\label{iv-1}
\ee
$$
=\frac{\ln n}{n}\Upsilon+\frac{1}{n}\varpi_4+o\left(\frac1n\right),
$$
where
\be
\Upsilon\equiv \frac{1}{12}\frac{F(0,0)}{\tilde{G}'_v(0,0)},\ \ \ \ \ \
\varpi_4\equiv -i\frac{2}{3}\frac{F(0,0)}{\tilde{G}'_v(0,0)}
\int_0^1 d\rho\ln\rho
e^{4i\rho\tilde{g}(0,0)}.
\label{alpha-4-5}
\ee
We have finally got the following expression for the integral $\Omega_n^{IV}$
\be
\Omega_n^{IV}=\frac{1}{n}\varpi_3+\frac{\ln n}{n}\Upsilon+\frac{1}{n}\varpi_4+o\left(\frac1n\right),
\label{iv-00}
\ee
where the constants  $\varpi_3$, $\Upsilon$ и $\varpi_4$ are given by the expressions 
(\ref{def-alpha3}), (\ref{alpha-4-5}).

\subsection{The Description of the Behaviour of the Contribution  $\Omega_n^{V}$.}

Pass on to the consideration of the last remaining contribution $\Omega_n^{V}$ into the normalization integral.
According to the definition (\ref{V-0})
$$
\Omega_n^{V}=\frac{1}{4n}\int_{\frac1n}^1\frac{du}{u}\int_0^1dve^{-4nuG(u,v)}\frac{\partial}{\partial v}
\left(\frac{F(0,v)}{G'_v(u,v)}\right).
$$
We introduce a new variable into the outer integral $t=nu$ and will integrate by parts in the inner integral:
$$
\Omega_n^{V}=\frac{1}{4n}\int_1^n\frac{dt}{t}\int_0^1dve^{-4tG(\frac{t}{n},v)}
\frac{\partial}{\partial v}
\left(\frac{F(0,v)}{G'_v(\frac{t}{n},v)}\right)=
$$
$$
=-\frac{1}{16n}\int_1^n\frac{dt}{t^2}\int_0^1dv \frac{\left(e^{-4tG(\frac{t}{n},v)}\right)'_v}
{G'_v(\frac{t}{n},v)}
\frac{\partial}{\partial v}
\left(\frac{F(0,v)}{G'_v(\frac{t}{n},v)}\right).
$$
Note that the following properties of the function $G(u,v)$:
$$
\Re G(u,v)\ge 0,\ \ \ \forall u\in[0,1],\ v\in[0,1],
$$
$$
G'_v\left(\frac{t}{n},v\right)=\beta+O\left(\sqrt{\frac{t}{n}}\right),\ \ \ \beta>0.
$$
Note as well that the obtained power law of the expression under the integral sign on the variable $t$, i.e. $t^{-2}$,
ensures the integral convergence on $dt$ at infinity and therefore an effective boundedness of the domain of integration
on $dt$.

From these considerations we obtain:
$$
\Omega_n^{V}=-\frac{1}{16n}\frac{1}{G'_v(0,1)}
\frac{\partial}{\partial v}\left(\frac{F(0,v)}{G'_v(0,v)}\right)\mathop{|}\limits_{v=1}
\int_1^\infty\frac{dt}{t^2}e^{-4tG(0,1)}+
\frac{1}{16n}\frac{1}{G'_v(0,0)}
\frac{\partial}{\partial v}\left(\frac{F(0,v)}{G'_v(0,v)}\right)\mathop{|}\limits_{v=0}+
$$
$$
+\frac{1}{16n}\int_1^\infty\frac{dt}{t^2}\int_0^1 dv e^{-4tG(0,v)}
\frac{\partial}{\partial v}\left[\frac{1}{G'_v(0,v)}\frac{\partial}{\partial v}
\left(\frac{F(0,v)}{G'_v(0,v)}\right)\right]+o(\frac1n).
$$
Finally, we come to the conclusion that 
\be
 \Omega_n^{V}=\frac{1}{n}\varpi_5+\frac{1}{n}\varpi_6+\frac{1}{n}\varpi_7+
o(\frac1n),
\label{def-V}
\ee
where the notations are used
\be
\varpi_5\equiv -\frac{1}{16}\frac{1}{G'_v(0,1)}
\frac{\partial}{\partial v}\left(\frac{F(0,v)}{G'_v(0,v)}\right)\mathop{|}\limits_{v=1}
\int_1^\infty\frac{dt}{t^2}e^{-4tG(0,1)},\ \ \ \ \ \ \
\varpi_6\equiv \frac{1}{16}\frac{1}{G'_v(0,0)}
\frac{\partial}{\partial v}\left(\frac{F(0,v)}{G'_v(0,v)}\right)\mathop{|}\limits_{v=0},
\label{alpha-56}
\ee
\be
\varpi_7\equiv \frac{1}{16}\int_1^\infty\frac{dt}{t^2}\int_0^1 dv e^{-4tG(0,v)}
\frac{\partial}{\partial v}\left[\frac{1}{G'_v(0,v)}\frac{\partial}{\partial v}
\left(\frac{F(0,v)}{G'_v(0,v)}\right)\right].
\label{alpha-7}
\ee

As a result we come to the conclusion that the asymptotics of the expression
$$
\Omega_n=\int_0^1 du \int_0^1 dv e^{-4n\beta uv f(u,v)}e^{4inu^{3/2}g(u,v)}F(u,v).
$$
as a large parameter $n\gg 1$ function is of the form:
\be
\Omega_n=\frac{1}{n}D_1(G)+\frac{\ln n}{n}D_2(G)+o\left(\frac1n\right),
\label{uv-110}
\ee
where the coefficients
\be
D_1(G)=\sum_{i=1}^7\varpi_i,\ \ \ \ \ \ D_2(G)=\Upsilon
\label{ast-n-0}
\ee
are defined in the expressions  (\ref{alpha-1}), (\ref{alpha-2}), (\ref{def-alpha3}), (\ref{alpha-4-5}),
(\ref{alpha-56}), (\ref{alpha-7}). The function $G$ is defined in the expression (\ref{def-g}).

\newpage

\section{Appendix D. The Asymptotics of the Scalar Product $\langle\tilde{\Psi}^{BBK},\psi_n^d\rangle|_{\bR^3_\bx}$.}

We will study here the asymptotics of the scalar product
$\langle\tilde{\Psi}^{BBK},\psi_n^d\rangle|_{\bR^3_\bx}(\by,\bq,\hk'')$
at large values of the variable $y\gg 1$ and large values of the index $n\gg 1$.

Introduce a notation
\be
Q\equiv \langle\tilde{\Psi}^{BBK},\psi_n^d\rangle|_{\bR^3_\bx}.
\label{def-q}
\ee
The expression ${\Psi}^{BBK}$ is defined above in the equations  (\ref{bbk0})-(\ref{phi-1}) for the domain 
$x\gg 1,\ \ y\gg 1$. We will remind that the expression $\tilde{\Psi}^{BBK}$ is defined for the extension 
${\Psi}^{BBK}$ in the domain of limit values of the variable $x$.
The expression  $\psi_n^d$ is defined in the equation  (\ref{psin}).

As was already discussed above in Section 6, we have a right to change the expression 
$\tilde{\Psi}^{BBK}$ for ${\Psi}^{BBK}$ in the integral (\ref{def-q}). Thus
\be
Q\equiv \langle{\Psi}^{BBK},\psi_n^d\rangle|_{\bR^3_\bx}+o\left(\frac1n\right).
\label{def-q1}
\ee
Use a weak asymptotics by $y\gg 1$ of the function ${\Psi}^{BBK}$ \cite{BL3}:
\be
\Psi^{BBK}\mathop{\sim}\limits_{y\rightarrow\infty}
B_0^{in}(\bq)\psi_c(\bx,\bk)\delta(\hy,-\hp)\frac{2\pi}{iyp}\left(1+i\frac{x}{y}
\langle\hx,\bL_{in}(\bq)\rangle\right)e^{-iyp+i\omega\ln y}-
\label{free}
\ee
$$
-B_0^{out}(\bq)\psi_c(\bx,\bk)\delta(\hy,\hp)\frac{2\pi}{iyp}\left(1+i\frac{x}{y}
\langle\hx,\bL_{out}(\bq)\rangle\right)e^{iyp+i\omega\ln y}.
$$
Here the notations are used 
$$
B_0^{in}(\bq)=A_0\Gamma(-i\eta_2)\Gamma(-i\eta_3)
e^{-\frac{\pi\omega}{2}}(1-e^{2\pi\eta_2})(1-e^{2\pi\eta_3})
\left[\frac{\sqrt{3}}{2}k_2(1-\langle\hat{\bp},\hat{\bk}_2\rangle)\right]^{i\eta_2}
\left[\frac{\sqrt{3}}{2}k_3(1+\langle\hat{\bp},\hat{\bk}_3\rangle)\right]^{i\eta_3},
$$
$$
\bL_{in}(\bq)=\frac{1}{\sqrt{3}}\left(\eta_2\frac{\hat{\bk}_2-
\hat{\bp}}{1-\langle\hat{\bp},\hat{\bk}_2\rangle}+
\eta_3\frac{\hat{\bk}_3+
\hat{\bp}}{1+\langle\hat{\bp},\hat{\bk}_3\rangle}\right),
$$
$$
B_0^{out}(\bq)=A_0\Gamma(-i\eta_2)\Gamma(-i\eta_3)
e^{-\frac{\pi\omega}{2}}(1-e^{2\pi\eta_2})(1-e^{2\pi\eta_3})
\left[\frac{\sqrt{3}}{2}k_2(1+\langle\hat{\bp},\hat{\bk}_2\rangle)\right]^{i\eta_2}
\left[\frac{\sqrt{3}}{2}k_3(1-\langle\hat{\bp},\hat{\bk}_3\rangle)\right]^{i\eta_3},
$$
$$
\bL_{out}(\bq)=\frac{1}{\sqrt{3}}\left(\eta_2\frac{\hat{\bk}_2+
\hat{\bp}}{1+\langle\hat{\bp},\hat{\bk}_2\rangle}+
\eta_3\frac{\hat{\bk}_3-
\hat{\bp}}{1-\langle\hat{\bp},\hat{\bk}_3\rangle}\right).
$$
Here the notations are used $\omega=\eta_2+\eta_3$,  $\ A_0=-\frac{1}{4\pi^2}N_0^{(23)}$,
where $\psi_c$ denotes a two-body scattering state.  The constant
$N_0^{(23)}=N_c^{(2)}N_c^{(3)}$ is expressed by means of the components 
$N_c^{(j)}=(2\pi)^{-\frac32}e^{-\frac{\pi\eta_j}{2}}\Gamma(1+i\eta_j)$.

Substituting the expression (\ref{free}) into the equation (\ref{def-q1}) 
and using the orthogonality of the functions
$\psi_c(\bx,\bk)$ и $\psi_n^d(\bx,\hk)$, we obtain 
\be
Q=iB_0^{in}(\bq)\frac{2\pi}{iyp}\delta(\hy,-\hp)\frac{1}{y}e^{-iyp+i\omega\ln y}
\int_{\bR^3}d\bx \langle\bx,\bL_{in}(\bq)\rangle\psi_c(\bx,\bk)\psi_n^d(\bx,\hk'')-
\label{q-def-1}
\ee
$$
-iB_0^{out}(\bq)\frac{2\pi}{iyp}\delta(\hy,\hp)\frac{1}{y}e^{iyp+i\omega\ln y}
\int_{\bR^3}d\bx \langle\bx,\bL_{out}(\bq)\rangle\psi_c(\bx,\bk)\psi_n^d(\bx,\hk'').
$$
To calculate the integrals in the expression (\ref{q-def-1}) we will now use again a weak asymptotics now
for the function $\psi_c(\bx,\bk)$. Here we again make an appeal to the fact that even though the function $\psi_n^d$ is
a two-body Coulomb operator discrete spectrum function, it actually belongs to the spectral vicinity of the discrete spectrum accumulation point of this operator.

Thus the decrease of the function at infinity in configuration space occurs very slowly (the power of decreasing 
exponent behaves as $\frac1n,\ \ n\gg 1$). In this sense the main contribution into the integrals in the 
expression (\ref{q-def-1}) is made by big values of the variable $x$, which allows to use weak asymptotics
\be
\psi_c(\bx,\bk)\sim N_c^{(1)}\left(\frac{e^{-ikx+i\eta\ln x}}{ikx}\delta(\hx,-\hk)-
\frac{e^{ikx+i\eta\ln x}}{ikx}s_c(\bx,\hk)\right),
\label{week-c}
\ee
where $s_c(\bx,\hk)$ is a two-body Coulomb scattering matrix. 

Define integral contributions in the converging and diverging waves in the equation (\ref{q-def-1}) in the following way:
\be
Z^{in}(\bq,\hk'')=\int_{\bR^3}d\bx \langle\bx,\bL_{in}(\bq)\rangle\psi_c(\bx,\bk)\psi_n^d(\bx,\hk''),
\label{zi-def}
\ee
\be
Z^{out}(\bq,\hk'')=\int_{\bR^3}d\bx \langle\bx,\bL_{out}(\bq)\rangle\psi_c(\bx,\bk)\psi_n^d(\bx,\hk'').
\label{zi-def-2}
\ee

Start with a calculation of the integral (\ref{zi-def}):
$$
Z^{in}(\bq,\hk'')=N_c^{(1)}\int_0^\infty x^3 dx\int_{\bS^2}d\hx\langle\hx,\bL_{in}(\bq)\rangle
e^{-\frac{|\alpha_1|}{2n}x}
L_{n-1}\left(\frac{|\alpha_1|}{2n}x(1-\langle\hk'',\hx\rangle)\right)\times
$$
$$
\times\left(\frac{e^{-ikx+i\eta\ln x}}{ikx}\delta(\hx,-\hk)-
\frac{e^{ikx+i\eta\ln x}}{ikx}s_c(\hx,\bk)\right).
$$
The expression in its turn is split into two terms to be defined as follows:
\be
Z^{in}=Z^{in}_1+Z^{in}_2,
\label{zi12-def}
\ee
where 
\be
Z^{in}_1\equiv \frac{N_c^{(1)}}{ik}\int_0^\infty x^{2+i\eta}
dx\int_{\bS^2}d\hx\langle\hx,\bL_{in}(\bq)\rangle
e^{-\frac{|\alpha_1|}{2n}x-ikx}
L_{n-1}\left(\frac{|\alpha_1|}{2n}x(1-\langle\hk'',\hx\rangle)\right)
\delta(\hx,-\hk),
\label{zi1-def}
\ee
\be
Z^{in}_2\equiv \frac{N_c^{(1)}}{ik}\int_0^\infty x^{2+i\eta}
dx\int_{\bS^2}d\hx\langle\hx,\bL_{in}(\bq)\rangle
e^{-\frac{|\alpha_1|}{2n}x+ikx}
L_{n-1}\left(\frac{|\alpha_1|}{2n}x(1-\langle\hk'',\hx\rangle)\right)
s_c(\hx,\bk).
\label{zi2-def}
\ee
Pass on directly to the calculation of the radial integral in the expression $Z^I_1$ (\ref{zi1-def}),
having preliminary integrated over the unit sphere:
$$
Z^{in}_1=\frac{N_c^{(1)}}{ik}\langle-\hk,\bL_{in}(\bq)\rangle
\int_0^\infty dx x^{2+i\eta} e^{-\frac{|\alpha_1|}{2n}x-ikx}
L_{n-1}\left(\frac{|\alpha_1|}{2n}x(1+\langle\hk'',\hk\rangle)\right).
$$
After the change of the variable $t=\frac{x}{n}$ we obtain
$$
Z^{in}_1=-C_n(\bq)\int_0^\infty dt t^{2+i\eta} e^{-\frac{|\alpha_1|}{2}t-iknt}
L_{n-1}\left(\frac{|\alpha_1|}{2}t(1+\langle\hk'',\hk\rangle)\right),
$$
where the notation is used 
$$
C_n(\bq)=n^{3+i\eta}\frac{N_c^{(1)}}{ik}\langle\hk,\bL_{in}(\bq)\rangle.
$$
According to  (7.414.7) \cite{Grad} the last integral is taken explicitly:
\be
Z^{in}_1=-C_n(\bq)\frac{\Gamma(3+i\eta)}{(\frac{|\alpha_1|}{2}+ikn)^{3+i\eta}}
{_2F_1}\left(1-n,3+i\eta;1;\frac{r}{\frac{|\alpha_1|}{2}+ikn}\right).
\label{zi1-00}
\ee
We have used here the notation $r=\frac{|\alpha_1|}{2}(1+\langle\hk'',\hk\rangle)$.

Make use of the following asymptotic presentation for the hypergeometric function
${_2F_1}$ [2.1.13] \cite{Bateman}:
$$
{_2F_1}(a,b;c;z)=\left(1+O\left(\frac{1}{b}\right)\right)\Phi(a,c,bz).
$$
Here  $c\neq 0,-1,-2,\dots$, $\ 0<|z|<1$, $\ |b|\rightarrow\infty$  so that 
$-\frac32\pi<\arg(bz)<\frac{\pi}{2}$.
In our case 
$$
a=3+i\eta,\ \ b=1-n,\ \ c=1,\ \ z=\frac{r}{\frac{|\alpha_1|}{2}+ikn}.
$$
In this sense the expression  (\ref{zi1-00}) takes form 
\be
Z^{in}_1=-\frac{N_c^{(1)}e^{\frac{\pi\eta}{2}}}{k^{4+i\eta}}\langle\hk,\bL_{in}(\bq)\rangle\Gamma(3+i\eta)
\Phi\left(3+i\eta,1,i\frac{|\alpha_1|}{2k}(1+\langle\hk'',\hk\rangle)\right).
\label{zi1-01}
\ee

The expression $Z^{in}_2$ (\ref{zi2-def}) can be obtained in the similar way when changing the order of integration. After integrating over  $dx$ we come to the expression
\be
Z^{in}_2=-\frac{N_c^{(1)}e^{-\frac{\pi\eta}{2}}}{k^{4+i\eta}}\Gamma(3+i\eta)
H^{in}(\bq,\hk''),
\label{zi2-01}
\ee
where the function $H(\bq,\hk'')$ is defined in the following way 
\be
H^{in}(\bq,\hk'')\equiv \int_{\bS^2}d\hx\langle\hx,\bL_{in}(\bq)\rangle
\Phi\left(3+i\eta,1,-i\frac{|\alpha_1|}{2k}(1-\langle\hk'',\hx\rangle)\right)
s_c(\hx,\bk).
\label{zi2-02}
\ee
Thus we have defined the integral contribution
 $Z^{in}(\bq,\hk'')$  (\ref{zi-def})
in the converging wave in the equation (\ref{q-def-1}) as a sum of the expressions described 
in the equations (\ref{zi1-01}) and (\ref{zi2-01})-(\ref{zi2-02}).

In the similar way we define an integral contribution $Z^{out}(\bq,\hk'')$  (\ref{zi-def-2}).

We have come to the conclusion that the expression
$\langle\tilde{\Psi}^{BBK},\psi_n^d\rangle|_{\bR^3_\bx}(\by,\bq,\hk'')$ (\ref{def-q})
is described in the following way 
\be
\langle\tilde{\Psi}^{BBK},\psi_n^d\rangle|_{\bR^3_\bx}(\by,\bq,\hk'')\sim
iB_0^{in}(\bq)\frac{2\pi}{iyp}\delta(\hy,-\hp)\frac{1}{y}e^{-iyp+i\omega\ln y}
Z^{in}(\bq,\hk'')-
\label{q-def-111}
\ee
$$
-iB_0^{out}(\bq)\frac{2\pi}{iyp}\delta(\hy,\hp)\frac{1}{y}e^{iyp+i\omega\ln y}
Z^{out}(\bq,\hk'').
$$
In accordance with the above said 
$$
Z^{in(out)}(\bq,\hk'')=-\frac{N_c^{(1)}\Gamma(3+i\eta)}{k^{4+i\eta}}\left[
e^{\frac{\pi\eta}{2}}\langle\hk,\bL_{in(out)}(\bq)\rangle
\Phi\left(3+i\eta,1,i\frac{|\alpha_1|}{2k}(1+\langle\hk'',\hk\rangle)\right)
-e^{-\frac{\pi\eta}{2}}H^{in(out)}(\bq,\hk'')\right],
$$
where 
$$
H^{in(out)}(\bq,\hk'')\equiv \int_{\bS^2}d\hx\langle\hx,\bL_{in,(out)}(\bq)\rangle
\Phi\left(3+i\eta,1,-i\frac{|\alpha_1|}{2k}(1-\langle\hk'',\hx\rangle)\right)
s_c(\hx,\bk).
$$

\end{document}